\documentclass[sn-nature]{sn-jnl} 

\usepackage{graphicx}%
\usepackage{multirow}%
\usepackage{amsmath,amssymb,amsfonts}%
\usepackage{amsthm}%
\usepackage{mathrsfs}%
\usepackage[title]{appendix}%
\usepackage{xcolor}%
\usepackage{textcomp}%
\usepackage{manyfoot}%
\usepackage{booktabs}%
\usepackage{algorithm}%
\usepackage{algorithmicx}%
\usepackage{algpseudocode}%
\usepackage{listings}%

\newcommand{\kms}{km\,s$^{-1}$}
\newcommand{\ms}{m\,s$^{-1}$}
\newcommand{\um}{{\textmu}m}

\newcommand{\water}{H$_{2}$O}
\newcommand{\cotwo}{CO$_{2}$}
\newcommand{\methane}{CH$_{4}$}
\newcommand{\qcotwo}{Q_{\rm CO_2}}
\newcommand{\qco}{Q_{\rm CO}}
\newcommand{\qmethane}{Q_{\rm CH_4}}

\begin{document}

\title[Chiron JWST DDT]{JWST reveals anomalously enhanced methane outgassing from below Chiron's water ice and carbon dioxide bearing surface}


\author*[1]{\fnm{Ian} \sur{Wong}}\email{iwong@stsci.edu}
\equalcont{These authors contributed equally to this work.}

\author[2]{\fnm{Silvia} \sur{Protopapa}}\email{silvia.protopapa@swri.edu}
\equalcont{These authors contributed equally to this work.}

\author[3]{\fnm{Aurélie} \sur{Guilbert-Lepoutre}}\email{aurelie.guilbert-lepoutre@univ-lyon1.fr}

\author[4]{\fnm{Geronimo L.} \sur{Villanueva}}\email{geronimo.l.villanueva@nasa.gov}

\author[1]{\fnm{Bryan} \sur{Holler}}\email{bholler@stsci.edu}

\author[5]{\fnm{Rosario} \sur{Brunetto}}\email{rosario.brunetto@universite-paris-saclay.fr}

\author[6]{\fnm{Joshua P.} \sur{Emery}}\email{joshua.emery@nau.edu}

\author[7,8]{\fnm{Noemí} \sur{Pinilla-Alonso}}\email{npinilla@uniovi.es}

\author[9]{\fnm{Ana Carolina} \sur{de Souza Feliciano}}\email{astro.carol@ucf.edu}

\author[9]{\fnm{Estela} \sur{Fernández-Valenzuela}}\email{estela@ucf.edu}

\affil*[1]{\orgname{Space Telescope Science Institute}, \orgaddress{\city{Baltimore}, \state{MD}, \country{USA}}}

\affil[2]{\orgname{Southwest Research Institute, Solar System Science and Exploration Division}, \orgaddress{\city{Boulder}, \state{CO}, \country{USA}}}

\affil[3]{\orgname{LGL TPE, UMR 5276 CNRS, Université Lyon 1, ENS}, \orgaddress{\city{Villeurbanne}, \country{France}}}

\affil[4]{\orgname{NASA Goddard Space Flight Center}, \orgaddress{\city{Greenbelt}, \state{MD}, \country{USA}}}

\affil[5]{\orgname{Université Paris-Saclay, CNRS, Institut d'Astrophysique Spatiale}, \orgaddress{\city{Orsay}, \country{France}}}

\affil[6]{\orgname{Northern Arizona University}, \orgaddress{\city{Flagstaff}, \state{AZ}, \country{USA}}}

\affil[7]{\orgname{Institute of Space Science and Technology of Asturias, University of Oviedo}, \orgaddress{\city{Oviedo}, \country{Spain}}}

\affil[8]{\orgname{Department of Physics, University of Oviedo}, \orgaddress{\city{Oviedo}, \country{Spain}}}

\affil[9]{\orgname{Florida Space Institute, University of Central Florida}, \orgaddress{\city{Orlando}, \state{FL}, \country{USA}}}


\abstract{Centaurs are inward-scattered Kuiper belt objects, with some exhibiting comet-like activity. The physical mechanisms powering this activity remain poorly understood, with carbon monoxide (CO) sublimation or the crystallization of amorphous water ice commonly invoked as the dominant drivers. Here we present high-resolution JWST spectroscopy of 2060 Chiron, one of the largest known Centaurs, revealing methane and carbon dioxide gas emission with distinct coma spatial morphologies and production rates of $\qmethane=(1.55\pm0.04)\times10^{27}$~molecules~s$^{-1}$ and $\qcotwo=(1.01\pm0.06)\times10^{26}$~molecules~s$^{-1}$. The surface spectrum displays spectral signatures attributed to water ice, carbon dioxide, CO, and refractory organic-rich material, while lacking detectable methane ice absorption bands. These findings suggest that carbon dioxide production is sustained by direct surface sublimation, whereas methane originates from the subsurface. The absence of measurable CO emission despite the presence of solid-state CO implies that any surviving primordial CO reservoir remains thermally inaccessible at greater depth below the methane, while irradiation-produced near-surface CO may be inefficiently released from the surface matrix. This inferred volatile stratification may result from long-term thermal evolution or potentially partial differentiation. Chiron differs markedly from other active small bodies, where CO production typically dominates over methane, indicating that Centaur activity may be driven by a broader range of volatile and thermophysical processes than predicted by canonical models.}


\maketitle

\section{Introduction}\label{sec:intro}

The Centaurs are a population of small bodies that originated in the Kuiper belt and were recently scattered onto eccentric orbits that traverse the giant planet region. Through close encounters with the giant planets, the orbits of Centaurs evolve chaotically, with short dynamical lifetimes on the order of 1--10~Myr, after which they are either ejected from the Solar System or scattered inward to become Jupiter-family comets \citep{levison1997,tiscareno2003,nesvorny2017}. For decades, Centaurs have served as valuable proxies for the more distant trans-Neptunian objects (TNOs). While recent improvements in telescope sensitivity and instrument capabilities have expanded the purview of observational exploration to encompass a wide swath of large- and mid-sized TNOs, detailed investigations of the more primitive $\sim$100~km sized objects remain daunting. By virtue of their smaller heliocentric distances, Centaurs are significantly more amenable to intensive spectroscopic study, with the potential to yield crucial insights into the chemical and physical nature of icy solar system planetesimals and the primordial conditions within the outermost regions of the protoplanetary disk. Meanwhile, the increased surface temperatures of Centaurs relative to TNOs also provide a unique window into the effects of secondary thermal processing on the surface properties of small icy bodies, potentially exposing subsurface materials that would otherwise remain inaccessible.

First reported in 1977 \citep{gehrels1977}, Chiron was the first Centaur to be discovered and has since become one of the most well-studied. Its 50.7-year orbit spans the region between Saturn and Uranus, with perihelion and aphelion distances of 8.5 and 18.9~AU, respectively. An effective diameter of $210\pm14$ km was derived from a combined analysis of thermal photometry at mid-infrared and radio wavelengths \citep{lellouch2017}, making Chiron the second largest known Centaur after Chariklo. The same study estimated an optical geometric albedo of $0.17\pm0.04$, which lies at the upper end of the measured range for the Centaur population \citep{romanishin2018}. At visible wavelengths, Chiron has a neutral spectral slope \citep{hartmann1990,campins1994}, similar to that of C-type asteroids. Early ground-based near-infrared spectroscopy detected weak absorption features at 1.5 and 2.0~\um, attributed to micron-sized water ice grains mixed with significant amounts of dust \citep{foster1999,luu2000}.

Chiron is remarkable for its recurrent cometary activity. Photometric monitoring over the past five decades has documented multiple outbursts, including one at the time of its discovery \citep{ortiz2015,dobson2024}. During the prolonged active phase that began in the late 1980s, when Chiron was inbound at 13~AU \citep{tholen1988}, and persisted through its perihelion passage in the mid-1990s, extensive observing campaigns captured a wealth of activity signatures. Ground-based imaging unveiled an extended, time-variable dust tail \citep{luu1990,meech1990}; stellar occultations indicated confined jets of solid material emanating from the surface and depositing debris in the near-nucleus environment \citep{elliot1995,bus1996}; near-infrared spectroscopy yielded the first detections of gaseous species in the coma, including cyanogen-bearing molecules \citep{bus1991} and carbon monoxide (CO) \citep{womack1999}. Subsequent stellar occultation campaigns have continued to probe Chiron's near-nucleus environment, revealing opaque structures that evolve on year-long timescales. While some features have been interpreted as concentric rings \citep{ortiz2015,ortiz2023,pereira2025}, others suggest more diffuse debris surrounding the object \citep{ruprecht2015,sickafoose2020,sickafoose2023,bragaribas2023}. 

Taken together, these observations highlight the complex and dynamic nature of Chiron. More broadly, active Centaurs have challenged traditional paradigms of cometary activity in the Solar System. Because these objects reside beyond the region where water ice readily sublimates, alternative mechanisms must be invoked to explain the observed outgassing. Current models converge on two primary processes driving Centaur activity: (a) direct sublimation of volatile ices (e.g., CO) from surface or subsurface layers \citep{cochran1991,womack2017,cabral2019,li2020,lilly2021}, and (b) the release of trapped volatiles during the crystallization of amorphous water ice \citep{prialnik1995,prialnik2008,guilbertlepoutre2012}. Detailed study of the comae and surface properties of active Centaurs promises to refine the understanding of distant cometary activity and, more generally, the formation and evolution of outer solar system planetesimals.

In March 2021, visible photometry of Chiron from the Asteroid Terrestrial-impact Last Alert System (ATLAS) uncovered a sharp increase in brightness, signaling another outburst and heralding a new epoch of activity \citep{dobson2021}. Occurring near Chiron's aphelion at 18.9~AU, this is the most distant report of Centaur activity by a significant margin and may indicate hitherto unattested mechanisms of cometary outgassing and surface evolution. JWST observations of Chiron acquired in July 2023 detected the presence of gaseous emission of methane (\methane) and carbon dioxide (\cotwo) and revealed prominent solid-state absorption features of water ice (\water), \cotwo, and CO \citep{pinillaalonso2024}. The follow-up JWST observations presented here were designed to provide substantially higher spectral resolving power and improved sensitivity across the 2--5~\um\ region, enabling line-resolved analyses of the gas fluorescence, quantitative production rate measurements, spatially resolved characterization of the gas comae, and a quantitative assessment of Chiron's surface composition. When interpreted within the broader context of active small bodies, the results of these analyses establish an unprecedented view of the coupling between Chiron's surface and volatile activity and provide critical insight into the physical and chemical processes behind Chiron's unique behavior.

\section{Results}\label{sec:res}

\subsection{JWST observations}\label{subsec:obs}

Chiron was observed on UT 2024 Jan 9 using the integral field unit (IFU) of the Near-Infrared Spectrograph (NIRSpec) \citep{jakobsen2022,boker2023} as part of JWST Cycle 2 Director's Discretionary Time proposal \#4621 (PI: Ian Wong). The NIRSpec IFU observing mode is ideal for studying active objects, as it provides the full spectrum at every pixel across the $3''\times3''$ field of view. At the time of the observations, Chiron was 18.72~AU from the Sun and 18.74~AU from JWST at a phase angle of 3.0$^{\circ}$. Two sets of four dithered exposures were obtained with the G235M and G395H gratings to produce spectra at resolving powers of $\sim$1000 and $\sim$2700, covering the wavelength ranges 1.7--3.1 and 2.9--5.1~\um, respectively. In order to facilitate detailed study of Chiron's coma, observations of a nearby background field using the same spectral settings were executed immediately following the on-target exposures. A full description of the observations and data processing methodology is provided in the Methods section.

Chiron's irradiance spectrum was extracted using a $0.5''\times0.5''$ square aperture centered on the target and converted into $I/F$ (Fig.~\ref{fig:1}, see Methods). The compounds responsible for the observed absorption and emission features are labeled in the figure. In addition to the notable fluorescence bands of gaseous \methane\ and \cotwo, there are clear detections of \water, \cotwo\ (including the $^{13}$C isotopologue), and CO ice absorption features, as well as the distinctive signature of aliphatic organics at 3.35--3.55~\um. In the following sections, the gas- and solid-phase properties of Chiron are analyzed in detail.

\subsection{Surface composition and solid-state modeling}\label{subsec:surface}

\begin{figure*}[ht]%
\centering
\includegraphics[width=\textwidth]{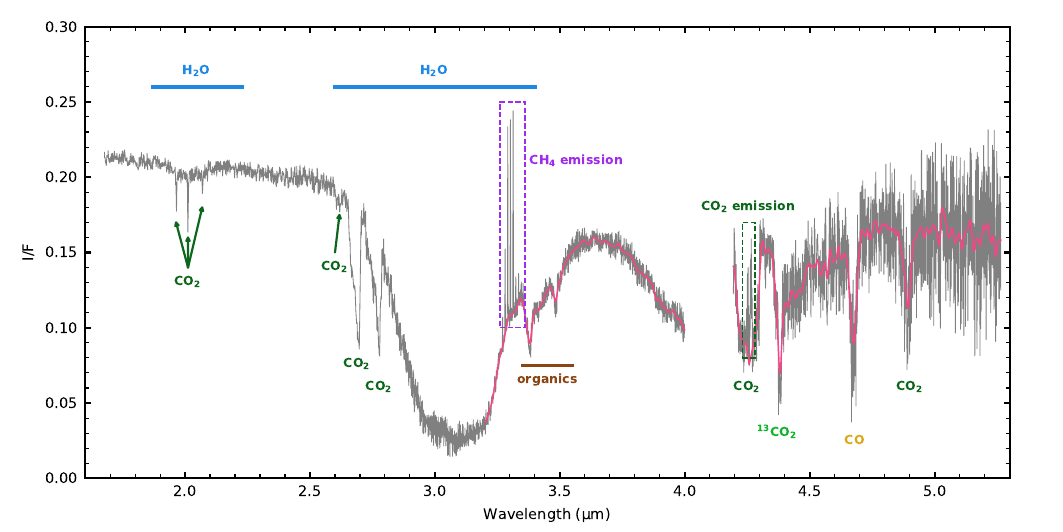}
\caption{\textbf{JWST/NIRSpec I/F spectrum of Chiron and detected solid-state and gaseous spectral features.} The measured spectrum, obtained using the G395H grating and extracted from a $5\times5$~pixel box centered on the centroid pixel, is shown in gray. Major molecular absorption and emission features are labeled. At wavelengths longer than 3.2~\um, a smoothed version of the spectrum with the \methane\ and \cotwo\ fluorescence bands removed is overplotted in magenta for clarity.}\label{fig:1}
\end{figure*}

Chiron exhibits an infrared spectrum that is remarkably rich in absorption features, revealing a diverse surface composition (Fig.~\ref{fig:1}). The spectrum displays multiple discrete absorption bands attributed to solid-state compounds, with well-defined band centers and shapes. In addition to prominent \water\ ice absorption bands at 2.0 and 3.0~\um, there are clear signatures of \cotwo\ ice, isotopically substituted carbon dioxide ($^{13}$\cotwo), organic materials, and CO. The radial profiles of Chiron's point-spread function (PSF), measured in both the 2.1--2.5~\um\ continuum region, which primarily probes refractory materials such as dust grains, and across the 2.65--2.80~\um\ \cotwo\ ice overtone bands, are indistinguishable from those of an inactive reference target (see Methods). In light of this, the identified spectral features are attributed to Chiron's surface rather than to icy grains ejected into the coma by activity. It is important to note, however, that the nondetection of a dust or ice coma does not necessarily preclude weak activity, as any associated solid component could remain below the detection limits of the PSF analysis.


Carbon dioxide is the spectroscopically dominant ice compound on Chiron, showing absorption bands across the full wavelength range of the observed spectrum. The detected spectral signatures include the narrow $2\nu_1+\nu_3$ (1.97~\um), $\nu_1+2\nu_2+\nu_3$ (2.01~\um), and $4\nu_2+\nu_3$ (2.07~\um) combination and overtone bands, as well as strong $\nu_1+\nu_3$ and $2\nu_2+\nu_3$ bands at 2.70 and 2.78~\um, respectively. The $\nu_3$ fundamental band at 4.27~\um\ is observed as a solid-state absorption with superposed \cotwo\ fluorescence emission (Fig.~\ref{fig:1}); the corresponding $\nu_3$ band of $^{13}$\cotwo\ is clearly detected at 4.38~\um. The $\nu_3$ fundamental band is characterized by a broad absorption bounded by two reflectance peaks that rise markedly above the surrounding continuum, closely resembling the distinctive ``double-dip'' \cotwo\ spectral profile identified on several TNOs and associated with spectrally prominent carbon oxides \citep{pinillaalonso2025,holler2025}. Although the short-wavelength side of this spectral feature is only partially sampled in the present dataset acquired with the NIRSpec G395H grating, previous observations obtained with the NIRSpec G395M grating provide full coverage of the 4--5~\um\ region and confirm the presence of the complete double-peaked \cotwo\ band morphology \citep{pinillaalonso2024}. This similarity in band structure suggests that the properties of Chiron's surface \cotwo\ are broadly consistent with those of its source population, i.e., TNOs, despite its subsequent dynamical evolution into the Centaur region. Other \cotwo\ absorption bands for which no specific vibrational modes have been assigned but which have been reported in laboratory measurements are also observed in the spectrum. In particular, absorption features at 2.62, 2.73, and 2.75~\um\ are detected, consistent with solid \cotwo\ absorption bands listed by Quirico \& Schmitt \citep{quirico1997}. Laboratory measurements by Hansen \cite{hansen1997} report an absorption band near 4.9~\um, which is clearly seen in Chiron's spectrum. However, this band is alternatively attributed to CO$_3$ and/or carbonyl sulfide (OCS) by Pinilla-Alonso~et~al. \cite{pinillaalonso2024} in their analysis of the July 2023 spectrum.

The simultaneous detection of multiple \cotwo\ bands across the full wavelength range further implies that \cotwo\ is present from near-surface layers down to depths of tens to hundreds of microns, as different transitions probe distinct optical path lengths within the ice \citep{protopapa2024}. The strength and multiplicity of the \cotwo\ absorptions on Chiron set it apart from other Centaurs and most TNOs, instead bearing a closer resemblance to the spectroscopic behavior observed on Triton \citep{cruikshank1993,quirico1999,grundy2010,holler2016,wong2025}. Nevertheless, the detectability of narrow absorption bands is strongly dependent on spectral resolving power. Convolution of Chiron's spectrum to the low spectral resolving power ($R=30-300$) of the prism mode, commonly used for NIRSpec observations of small solar system bodies, renders the short-wavelength \cotwo\ bands in the 2.0-\um\ region indiscernible. This is consistent with prism-mode spectra of TNOs and Centaurs that display the \cotwo\ features at 2.7 and 4.27~\um\ but lack the 2.0-\um\ triplet. However, spectral resolution alone cannot account for differences in spectral behavior observed in high-resolution data. Both Chiron and Charon have been observed with NIRSpec at $R\sim2700$, yet their spectra differ markedly: while Charon exhibits the $\nu_1+\nu_3$ and $\nu_3$ bands near 2.7 and 4.3~\um, it lacks the additional overtone and combination bands at shorter wavelengths present in Chiron's spectrum \citep{protopapa2024}. This contrast indicates that the presence of the \cotwo\ bands near 2.0~\um\ is not solely a function of resolving power, but instead reflects intrinsic differences in \cotwo\ abundance and/or its vertical distribution within the surface. In particular, Chiron's spectrum suggests either a higher overall \cotwo\ abundance or a distribution that allows photons to sample deeper layers where these weaker transitions become detectable.

To interpret the solid-state \water\ and \cotwo\ absorption features, Chiron's 1.8--3.2~\um\ spectrum---encompassing most \cotwo\ absorption bands except the $\nu_3$ fundamental—was modeled using the Hapke radiative transfer formalism \citep{hapke1993,hapke2012}. The best-fit model configuration is an intimate mixture of three components: (1) an aggregate of crystalline \water\ ice and \cotwo, comprising 4.5\% of the mixture with an effective path length of 36~\um; (2) amorphous \water\ ice, accounting for 0.2\% with grain sizes of 1.3~\um; and (3) tholin-like materials, which dominate the mixture with a volume fraction of 95.3\% and represented by grains with an effective path length of $\sim$500~\um.  Within the aggregate, \water\ ice accounts for 2.6\% of the mixture, with the remaining fraction attributed to \cotwo. Several configurations were explored, including areal mixtures and layered geometries; among these, a layered configuration with components (1) and (2) overlying component (3) provided a statistically equivalent fit to the intimate mixture model described above. A comparison between the data and the best-fit model across the 1.8--3.2~\um\ wavelength range is shown in Fig.~\ref{fig:2}a,b. The modeling successfully reproduces both the strength and shape of the absorption bands, as well as the overall $I/F$ level and continuum profile. The root mean square deviation of the best-fit model from the measured spectrum is only 1\% higher than the average scatter of the data across the modeled wavelength range, which is quantified as the standard deviation of the residual array. For details on the modeling methodology, see Methods. 

The broad absorption band near 2.0~\um, together with the feature at 1.5~\um\ that lies outside the wavelength coverage of the spectrum presented here, has been attributed in the literature to \water\ ice overtone and combination bands, with possible contributions from –OH, –NH, and –CH functional groups in complex organic materials \citep{pinillaalonso2024}. Within the wavelength interval probed, however, the best-fit spectral model shown in Fig.~\ref{fig:2}a,b reproduces the observed band strengths and shapes without requiring additional absorptions from –OH, –NH, or –CH groups beyond those associated with \water\ ice. Moreover, the tholin-like materials used in the best-fit model, representative of Titan-type tholins, do not intrinsically exhibit these functional group absorptions in this spectral region. 

In addition to the \water\ and \cotwo\ features discussed above, Chiron displays a prominent absorption band complex between 3.3 and 3.6~\um, indicative of organic compounds. Pinilla-Alonso~et~al. \cite{pinillaalonso2024} attributed this complex structure to C--H stretching and combination modes of aliphatic hydrocarbons, identifying ethane (C$_2$H$_6$) and propane (C$_3$H$_8$) as the most likely contributors, with acetylene (C$_2$H$_2$) proposed as the carrier of the 3.1-\um\ feature. To investigate the nature of Chiron's organic features, particularly the absorption complex spanning 3.3--3.6~\um, model spectra were generated for C$_2$H$_2$, C$_2$H$_6$, C$_3$H$_4$, C$_3$H$_8$, and ice tholins using laboratory-measured optical constants (see Methods); these are plotted alongside Chiron's spectrum in Fig.~\ref{fig:2}c. While ice tholins, produced by irradiation of \water-dominated mixtures containing light hydrocarbons (\water/C$_2$H$_6$ $\approx$ 6:1) \citep{khare1993}, match the overall breadth of the absorption complex, they do not capture the individual narrow bands present in Chiron's spectrum. Among the individual molecular candidates, C$_3$H$_8$ provides the closest match to both the positions and shape of the observed narrow bands, whereas the other hydrocarbons either fail to reproduce the band centers or introduce additional features that are not observed. Therefore, the comparison favors C$_3$H$_8$, perhaps in conjunction with ice tholins, as the leading carrier of the 3.3--3.6~\um\ absorption feature on Chiron. A weak absorption near 3.07~\um\ may be present in the data at the level of the noise; however, reproducing this feature does not statistically require the inclusion of C$_2$H$_2$ (Fig.~\ref{fig:2}c). In the modeling, the absorption at $\sim$3.1~\um\ is instead due to crystalline \water\ ice (Fig.~\ref{fig:2}a).

\subsection{Coma modeling and mapping}\label{subsec:coma}

At the spectral resolution of the G395H grating, the individual fluorescence bands of \methane\ and \cotwo\ are clearly resolved, enabling the first quantitative assessment of the respective production rates. Chiron's outgassing was characterized by modeling the fluorescence of each molecule with the Planetary Spectrum Generator (PSG) \citep{villanueva2018,villanueva2022}; see Methods for a full description of the procedure. After subtracting the local continuum from the irradiance spectrum, the residual emission was fitted with an isothermal, expanding coma model to simultaneously retrieve the gas production rate ($Q$) and the coma rotational temperature. To facilitate direct comparison between the \methane\ and \cotwo\ comae, the expansion velocity at Chiron's heliocentric distance was fixed to the value given by empirical scaling law of Delsemme \cite{delsemme1982}---130~\ms.

\begin{figure*}[ht]%
\centering
\includegraphics[width=\textwidth]{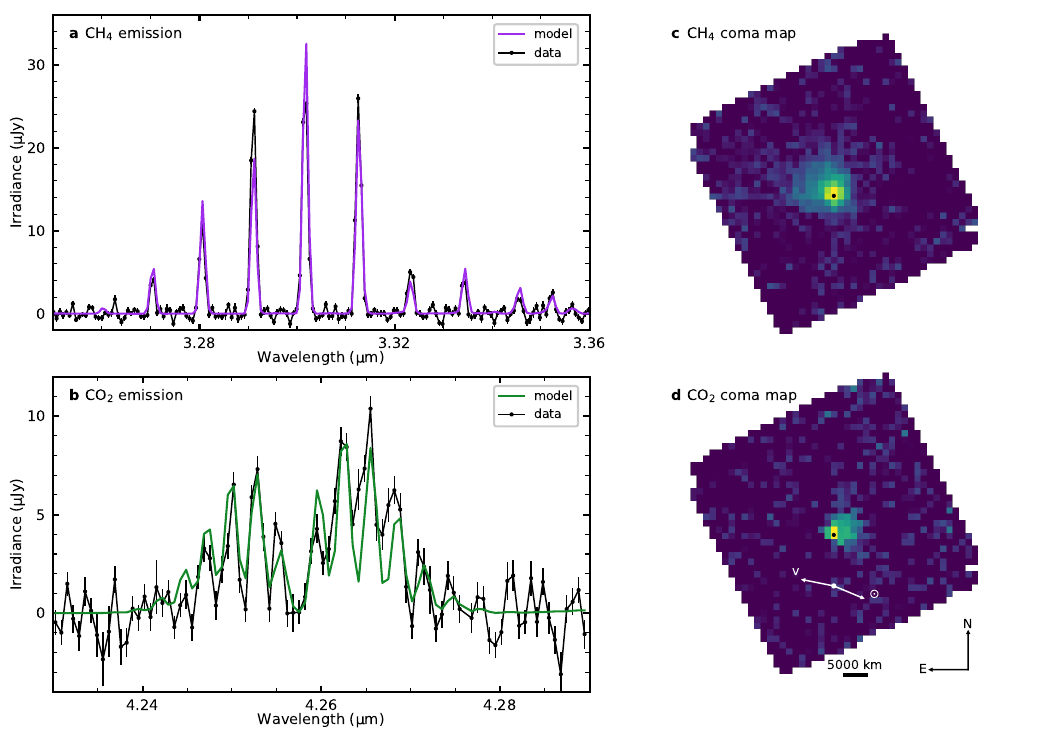}
\caption{\textbf{Modeling and mapping of Chiron's coma.} \textbf{a,b} Continuum-subtracted irradiance spectrum of Chiron in the \methane\ and \cotwo\ fluorescence regions (black), extracted from a $5\times5$~pixel box around the centroid pixel, with the best-fit PSG non-LTE coma models overplotted in purple and green, respectively. The production rate and rovibrational temperature derived from the PSG retrievals are shown in each panel, assuming an expansion velocity of 130~\ms. \textbf{c,d} Spatial maps of the \methane\ and \cotwo\ emission, calculated from a pixel-by-pixel band area analysis, showing the strongly discrepant coma morphologies. The sunward and sky-projected velocity directions at the time of the observations are indicated.}\label{fig:3}
\end{figure*}

The best-fit gas production rates and uncertainties for \methane\ and \cotwo\ are $\qmethane = (1.55\pm0.04) \times 10^{27}$ and $\qcotwo = (1.01\pm0.06) \times 10^{26}$ molecules per second, respectively. The retrieved coma temperatures for the two species are comparable---$24\pm1$~K for \methane\ and $18\pm2$~K for \cotwo. The continuum-subtracted spectra in the \methane\ and \cotwo\ fluorescence regions, together with the corresponding best-fit PSG coma models, are presented in Fig.~\ref{fig:3}a,b. In the case of \methane, nine sharp rotational lines are fully resolved between 3.26 and 3.36~\um. The coma model reproduces the observed line positions and peak amplitudes with high fidelity. However, the discrepancy in the relative strengths of the three strongest fluorescence peaks may indicate nonuniformity in the gas rotational temperature throughout the coma. For \cotwo, while the narrow lines are partially blended at the spectral resolution of the data, ten distinct fluorescence peaks within the P- and R-branch structure are discernible above the noise level. 

A careful inspection of Chiron's irradiance spectrum did not uncover evidence of other fluorescing gas molecules in the coma. Particularly notable is the absence of identifiable CO emission near 4.7~\um, where the molecule's primary rotational lines reside; instead, the spectrum displays a strong CO-ice absorption feature centered at 4.67~\um. However, due to the comparatively low signal-to-noise ratio of the data at those wavelengths and the relatively weak intrinsic line strengths of CO, a modest level of CO outgassing cannot be excluded. An analogous analysis of the irradiance spectrum between 4.60 and 4.75~\um\ yielded a $1\sigma$ upper limit of $Q_{\rm CO} < 1.8 \times 10^{25}$ molecules per second. For comparison, millimeter-wavelength spectroscopy obtained shortly before Chiron's perihelion passage in 1995 measured a CO production rate of $1.5 \times 10^{28}$ molecules per second \citep{womack1999}. 

The NIRSpec IFU data reveal the spatial distribution of the outgassed molecules in the vicinity of the nucleus. The extent of \methane\ and \cotwo\ across the field of view was derived by measuring the integrated band area of each species' rotational lines on a pixel-by-pixel basis (see Methods). The resultant coma maps are displayed in Fig.~\ref{fig:3}c,d, along with a scale bar and vectors indicating the direction of Chiron's motion and incident solar irradiation at the time of the observations. The key finding is that the \methane\ and \cotwo\ comae show markedly different spatial distributions, possibly indicating that they originate from distinct regions on the surface. \methane\ is present in a broad fan-shaped tail extending more than 10,000~km to the northeast of the central body. In contrast, the \cotwo\ coma is more compact, emerging to the west and spanning roughly 5,000~km. 

In the case of \cotwo, the production region can be constrained from thermodynamical considerations. Chiron was near aphelion, at a heliocentric distance of 18.72~AU, during the JWST observations. Assuming the measured geometric albedo from \cite{fornasier2013}---$p_v = 0.16$---an emissivity of $\epsilon=0.9$, and a beaming factor of $\eta = 0.756$, the average temperature of the sunward-facing hemisphere was 69~K, below the sublimation temperature of \cotwo\ ($\sim$80~K; \cite{prialnik2004}). However, the temperature at the sub-solar point was significantly higher---around 100~K. This suggests that \cotwo\ production on Chiron during the 2024 apparition was likely concentrated near the sub-solar region.

\section{Discussion}\label{sec;disc}

The combined insights from the spectral, thermodynamical, and coma morphology analyses point to distinct source regions for the observed gaseous species. Given the presence of \cotwo-bearing aggregates on the surface and the thermal conditions prevailing on Chiron during the 2024 observations, direct \cotwo\ sublimation is viable in the warmest surface regions, most plausibly near the sub-solar point. This supports a surficial origin for the observed \cotwo\ gas without the need to invoke the amorphous-to-crystalline phase transition of water ice, which has often been posited as a primary activity trigger on Centaurs \cite{prialnik1995,jewitt2009}. Moreover, the surface temperatures on Chiron near aphelion are too cold to readily activate the crystallization process, which is expected to occur at temperatures between 110 and 150~K on timescales of minutes to years \cite{protopapa2021}. Meanwhile, the simultaneous presence of \methane\ gas emission and absence of detectable \methane\ ice absorption features in Chiron's spectrum (Fig.~\ref{fig:1}) suggests that the observed \methane\ likely originates from subsurface volatile reservoirs that are not directly detectable in reflectance. Pinilla-Alonso~et~al. \cite{pinillaalonso2024} alternatively proposed that \methane\ trapped within amorphous \water\ ice at or near the surface could be released through thermally driven desorption associated with structural evolution of the ice.

\begin{figure*}[ht]%
\centering
\includegraphics[width=0.8\textwidth]{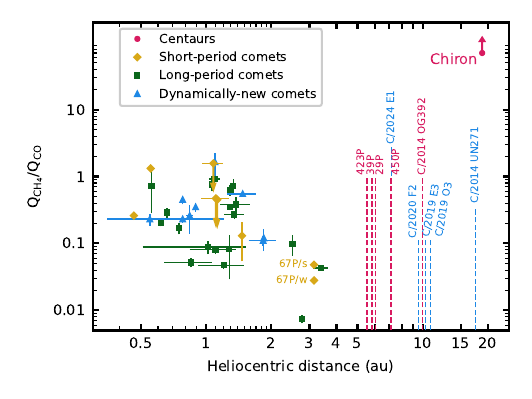}
\caption{\textbf{\methane-to-CO abundance ratio measured in Chiron's coma in the context of other active objects in the Solar System.} A compilation of published \methane:CO production rate ratios \cite{leroy2015,dellorusso2016,bonev2017,disanti2017,roth2017,disanti2018,faggi2018,faggi2019,mckay2019,disanti2021,faggi2021,dellorusso2022,faggi2023} as a function of heliocentric distance at the time of observation for Centaurs (magenta points), short-period comets (yellow diamonds), long-period comets (green squares), and dynamically-new comets (blue triangles). Data points with arrows indicate upper or lower $1\sigma$ limits. For 67P/Churyumov-Gerasimenko, the measurements from the summer and winter hemispheres are shown separately. Chiron presents an anomalously high $\qmethane/\qco$ ratio. The vertical dashed lines denote objects with extant JWST observations, but for which abundance ratios have not been reported.}\label{fig:4}
\end{figure*}

The most notable finding from the JWST observations is the simultaneous presence of \methane\ production and absence of detectable CO emission, establishing Chiron as an anomaly within the broader context of active objects within the Solar System. Figure~\ref{fig:4} shows a compilation of measured $\qmethane/\qco$ ratios from the literature \cite{leroy2015,dellorusso2016,bonev2017,disanti2017,roth2017,disanti2018,faggi2018,faggi2019,mckay2019,disanti2021,faggi2021,dellorusso2022,faggi2023}. The upper limit on the CO production rate derived from the Chiron observations corresponds to a $1\sigma$ lower limit of $\qmethane/\qco = 70$, indicating that Chiron's coma is enhanced in \methane\ (or, alternatively, depleted in CO) by at least two orders of magnitude relative to typical short-period, long-period, and dynamically-new comets, which have median and 75th percentile $\qmethane/\qco$ values of 0.26 and 0.49, respectively. Among active Centaurs, Chiron is the only object for which \methane\ emission has been detected. Recent JWST observations of 29P \cite{faggi2024}, 39P \cite{harringtonpinto2023}, and C/2024~E1 \cite{snodgrass2025} revealed only \cotwo\ and/or CO, with no evidence of \methane. However, formal upper limits on the \methane\ production rate were not reported for those objects.

The detection of \methane\ emission in the absence of measurable CO is particularly intriguing from the standpoint of volatile evolution. Both \methane\ and CO are hypervolatile species that sublimate readily even at the low surface temperatures characteristic of the Kuiper belt \cite{schaller2007,fray2009,protopapa2025}. Nevertheless, they have been observed in the comae of kilometer-scale comets, implying that their long-term survival in the interior of small bodies is not, by itself, unexpected. However, when both species are detected, CO is typically observed at higher production rates than \methane\ due to its higher volatility. Therefore, the lack of detectable CO during Chiron's aphelion apparition suggests a degree of vertical stratification between the accessible \methane\ and CO reservoirs within the subsurface. 

One possible scenario arises from the difference in volatility between \methane\ and CO. Heating of the surface leads to progressive depletion of the primordially accreted reservoirs of hypervolatile species from the outermost layers, with the sublimation front of CO receding much faster into the interior than that of \methane. This process would result in a vertically stratified configuration in which the surviving reservoir of \methane\ lies closer to the surface than CO and is consequently more accessible to activation from surface insolation. It follows that Chiron at aphelion may have been in a particular thermophysical state in which the thermal wave from incident sunlight penetrated only to the shallower \methane-bearing layer, while the CO reservoir at greater depth remained dormant. Such volatile stratification is expected in thermally evolved Centaurs, where long-term orbital evolution progressively alters the structure and composition of the subsurface \cite{guilbertlepoutre2023}. Extending this argument to the wider ensemble of active objects, the relative \methane-to-CO production rate is strongly controlled by the level of radiative forcing and is therefore expected to vary with heliocentric distance. For the closest-in comets, with perihelia within 3~au, such as the majority of objects plotted in Fig.~\ref{fig:4}, the intense insolation drives the thermal wave deep into the interior, fully activating both the \methane\ and the CO ice reservoirs. In this regime, the resulting production rates are governed by the relative volatility of the two species, yielding a CO enrichment in the comae of those objects.

However, the behavior of other active Centaurs presents an apparent conundrum. Most active Centaurs, including 29P and 39P, orbit closer to the Sun than Chiron, yet remain significantly less heated than low-perihelion comets (e.g., 67P). Within this thermodynamical regime, one would expect the production of \methane\ and CO to be highly sensitive to the aforementioned vertical stratification of the sublimation fronts, with \methane\ being the more accessible species due to its shallower depletion depth. Nevertheless, while CO has been detected in the comae of several active Centaurs, \methane\ has so far not been reported in any Centaur other than Chiron. 

A plausible resolution to this quandary may be that the CO observed in the comae of other active Centaurs does not originate from deep primordial reservoirs, which may remain thermally inaccessible, but instead derives from secondary near-surface repositories produced through long-term irradiation of carbon-bearing ices. Laboratory studies have demonstrated that CO is a byproduct of \cotwo\ and CH$_3$OH radiolysis \cite{Brucato1997,Mejia2015,Quirico2023,Henault2025}. Prolonged irradiation of surfaces rich in carbon-bearing ices may therefore lead to enrichment of CO within the near-surface layers, potentially contributing to the observed CO absorption in Chiron's reflectance spectrum (Fig.~\ref{fig:1}). The resulting CO may remain trapped within the irradiated matrix and diffuse inefficiently even as \cotwo\ sublimates. Within this framework, the higher levels of solar insolation experienced by closer-in active Centaurs may promote the release of both \cotwo\ and irradiation-produced near-surface CO. Chiron, by contrast, may currently reside in a colder thermophysical regime in which \cotwo\ sublimation can occur while much of the irradiation-generated CO remains trapped within the surface matrix, therefore explaining its observed \methane-rich and CO-poor coma composition.

An alternative scenario that could account for Chiron's unusual \methane-dominated gas production is related to its large size. At 210~km in diameter, Chiron is significantly larger than most active objects, which are typically only a few kilometers to a few tens of kilometers across. Its larger size and potentially greater abundance of radioactive isotopes may have promoted enhanced internal heating early in its evolution, possibly leading to partial differentiation and large-scale volatile migration within its interior. Such internal processing could have yielded a vertically stratified volatile distribution in which the near-surface layers are enriched in \methane, with CO-bearing material sequestered at greater depths. 

Ultimately, distant Centaur activity likely reflects a complex interplay between multiple physical mechanisms, including radiative forcing, irradiation-driven volatile processing, and dynamical history. Additional observations of active objects will be essential for disentangling these various mechanisms. Over the past four years, spectroscopic measurements of a diverse sample of active Centaurs and dynamically-new comets spanning a broad range of heliocentric distances and object diameters have been obtained with JWST. The heliocentric distances corresponding to these observations are indicated by the vertical dashed lines in Fig.~\ref{fig:4}. Comparative analyses of uniformly derived production rates and upper limits for \methane, \cotwo, and CO across the full sample of active objects will help reveal systematic trends in volatile outgassing and elucidate the dominant drivers of activity in distant icy bodies.


\section{Methods}\label{sec:methods}

\subsection{Observations and data processing}\label{subsec:proc}

Observations of Chiron were executed on UT 2024 Jan 9 between 02:31 and 03:54 using the NIRSpec IFU. Exposures were collected with two grating--filter combinations---G235M/F170LP and G395H/F290LP---covering the wavelength ranges 1.66--3.07 and 2.87--5.14~\um\ at spectral resolving powers of $\sim$1000 and $\sim$2700, respectively. The higher spectral resolution setting disperses the spectral trace across two NIRSpec detectors, resulting in a gap in wavelength coverage spanning 4.0--4.2~\um. A 4-point dither pattern was employed, with individual effective exposure times of 277 and 598 seconds in the two grating settings, yielding total exposure times of 1108 and 2392 seconds, respectively. To reduce the level of correlated noise during detector readout, the NRSIRS2RAPID readout mode was selected \citep{moseley2010,rauscher2012}. A pair of dithered exposures of an empty background region near the target was obtained in each grating setting immediately following the on-target observations, using the same integration times and readout mode.

The overall data reduction workflow closely matched the methodology used in previous analyses of NIRSpec observations of solar system objects \citep{emery2024,grundy2024,wong2024}. The uncalibrated files were processed locally using Version~1.13.4 of the official JWST calibration pipeline \citep{jwst}, with reference files drawn from context \texttt{jwst\_1214.pmap} of the JWST Calibration Reference Data System. Both the target and the background images were passed through Stage 1 of the pipeline to produce bias- and dark-corrected count rate images. Some residual readnoise was present across these images, resulting in systematic flux level offsets between adjacent detector columns. Stage 2 of the calibration pipeline includes an optional additional readnoise correction subroutine based on the NSClean method \citep{rauscher2024}. The count rate images of the target and background were passed through the first three steps of the Stage 2 pipeline processing, ending with the readnoise correction step, which was manually turned on. The resulting readnoise-corrected images were significantly cleaner, with no discernible vertical banding.

The sky background was computed using the dedicated background step contained within Stage 2 of the calibration pipeline, which combined the pairs of dithered background exposures via a pixel-by-pixel mean. After subtracting the background, the corrected Chiron count rate images were passed through the full Stage 2 pipeline processing to produce spatially-rectified, distortion-corrected, wavelength- and flux-calibrated data cubes for each dithered exposure. Each slice within the data cubes corresponds to a $3''\times3''$ image of the IFU field of view at a single wavelength within the fixed wavelength grid, projected to the equatorial coordinate system with a pixel scale of $0.1''$.

To create the final dither-combined data cubes in each grating setting for use in spatial analyses of Chiron's coma, the cal.fits files were processed through Stage 3 of the JWST pipeline. For moving targets, the pipeline takes into account the target's motion and the relative dither offsets to project all input exposures into the target's co-moving frame. The Stage 3 processing includes an outlier detection step that can sometimes erroneously flag pixels near the centroid, particularly in the case of moving, extended sources. Following the recommendation outlined in the JWST Documentation, this outlier detection step was manually turned off when processing the Chiron observations. Instead, outlier detection in the dither-combined data cubes was handled using custom routines during the coma mapping analysis.

The default flux units output by the JWST pipeline are surface brightness units (MJy~sr$^{-1}$). Prior to spectral extraction, the flux values were converted to irradiance units (MJy) using the pixel area listed in the file headers.

\subsection{Spectral extraction}\label{subsec:extr}

The centroid of Chiron's point-spread function (PSF) was calculated by median-averaging the data cube along the wavelength axis and fitting a 2D Gaussian. All pixels with a nonzero data quality flag value were masked. Remaining outliers were flagged using two different methods. Across the outer region, which was defined as all pixels outside of the $5\times5$ pixel box centered on the centroid, $5\sigma$ outliers were iteratively masked slice by slice until none remained. For the central region within the $5\times5$ pixel box, outliers were identified by extracting each pixel's spectrum across all wavelength slices, fitting the spectrum with a cubic spline, and masking points that deviated from the smoothed trend line by more than $10\sigma$. The higher outlier threshold here was chosen to prevent the routine from erroneously flagging the high flux values within the gas emission peaks. 

Spectral extraction was then performed on the individual dithered exposures within the $5\times5$ pixel box centered on the centroid. The flux uncertainties were calculated as the quadrature sum of the individual pixel flux errors within the extraction region. To remove outlier points in the spectra while preserving narrow astrophysical signals, each individual dither spectrum was first processed with a 21-point-wide moving median filter to mask $10\sigma$ outliers. Next, the spectra were plotted together and carefully inspected against one another to facilitate manual masking of lower-level outlier points in the individual dither spectra. Lastly, the dither spectra were mean-averaged to produce the final irradiance spectrum. The corresponding flux uncertainties were computed using a quadrature sum, divided by the number of unmasked points across the dither spectra at each wavelength.

The JWST pipeline includes an automatic Doppler correction for the component of the telescope's barycentric velocity in the pointing direction, which is listed in the science header of the data cubes as VELOSYS. However, in the case of solar system objects, the orbital motion of the target is not reliably accounted for (as of Version~1.13.4). To properly recast the wavelength solution into the stationary frame of the target without the need to interpolate and resample the individual dither spectra, the pipeline's Doppler correction was first removed using the average VELOSYS value across the JWST observations (28.86~\kms). Then, the wavelength grids were Doppler shifted using the true radial velocity of Chiron relative to JWST, as provided by the JPL Horizons ephemerides service at the midpoint time of the observations (29.47~\kms). The dispersion in VELOSYS values and Chiron's instantaneous relative velocity across the individual exposures is less than 0.01~\kms; the corresponding spread in Doppler wavelength shifts is negligible in comparison to the wavelength grid spacings.

The full irradiance spectrum of Chiron is plotted in Fig.~\ref{fig:sm1}a. Chiron has a well-measured rotational period of $5.917813\pm0.000007$~hr and a small rotational brightness variation of $<0.1$~mag \citep{marcialis1993,ortiz2015}, indicating that the object does not deviate significantly from a spherical shape. While the JWST observations of Chiron spanned roughly a quarter of the rotational period, no systematic offset was found between the overlapping wavelengths of the extracted G235M and G395H irradiance spectra, and therefore the spectra were not renormalized prior to further analysis.

To derive the reflectance spectrum, publicly-accessible NIRSpec observations of the G-type solar analog star GSPC~P330-E (herefter, P330-E) from Program \#1538 (PI: Karl Gordon) were used. The irradiance spectra of the standard star in the G235M and G395H grating settings were extracted using an analogous methodology to the processing of the Chiron observations. The only substantive difference was the lack of dedicated background exposures for the P330-E observations. To remove the background flux level from each wavelength slice, the median value of all pixels outside of a $21\times21$ pixel box centered on the centroid was subtracted prior to summing the source flux within the $5\times5$ pixel extraction region. The combined irradiance spectrum of P330-E was Doppler shifted to account for the $-$53~\kms\ barycentric radial velocity of the target \citep{soubiran2018}; a scaled version of this spectrum is plotted in Fig.~\ref{fig:sm1}a. Dividing the star's spectrum from that of Chiron produced the reflectance spectrum, while self-consistently correcting for any common-mode instrumental systematics shared by both spectra.

\subsection{PSF analysis}\label{subsec:psf}

To search for excess solid material surrounding the main body, the PSF of Chiron in the dither-combined data cubes was compared to the PSFs of other point-source targets observed with the NIRSpec IFU. The data cubes from the G235M observations were the focus of this analysis, due to the higher signal-to-noise ratio of the source PSFs at those wavelengths. NIRSpec PSFs have a complex spatial profile, including a sharp central peak and six lobe-shaped secondary diffraction maxima at larger separations---a consequence of the hexagonal shape of JWST's mirrors. Furthermore, the PSF shape is severely undersampled at the $0.1''$ pixel scale of the IFU wavelength slices. It follows that the precise location of a source's centroid relative to the pixel boundaries greatly affects the shape of the measured radial profile of the PSF at small separations ($<1.5$~pixels). For a detailed comparison with Chiron, NIRSpec data of the inactive Centaur 2013~XZ8 were reduced following an analogous data reduction procedure. The publicly-accessible observations of 2013~XZ8 were obtained as part of Cycle 1 Guaranteed Time Observations Program \#1272 (PI: Dean Hines). This target was selected due to its nearly identical relative intrapixel centroid position.

\begin{figure*}[ht]%
\centering
\includegraphics[width=0.8\textwidth]{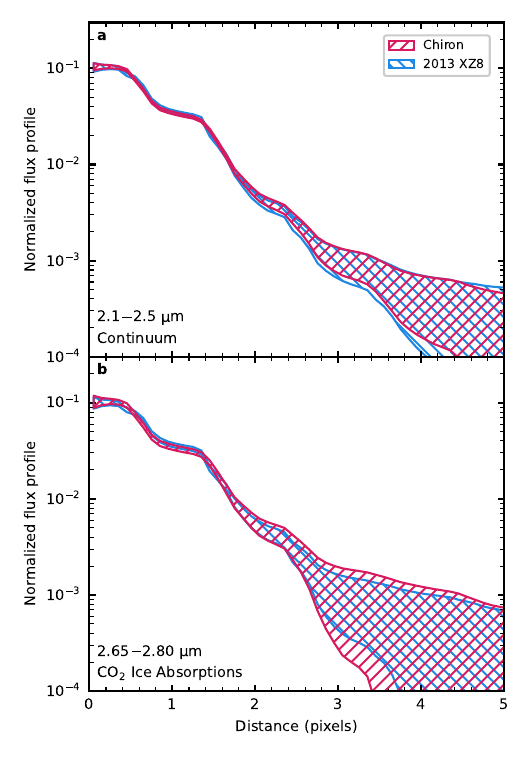}
\caption{\textbf{Probing for extended emission around Chiron.} \textbf{a} The $1\sigma$ bounded region of Chiron's azimuthally averaged radial flux profile (magenta), measured across the 2.1--2.5~\um\ continuum region and normalized to an integrated flux of unity, alongside the corresponding radial flux profile of the inactive Centaur 2013~XZ8 (cyan). \textbf{b} Same as panel \textbf{a}, but for the 2.65--2.80~\um\ wavelength range, which includes the combination bands of \cotwo\ ice. In both wavelength ranges, Chiron's PSF is indistinguishable from that of a point source.}\label{fig:sm2}
\end{figure*}

The radial profiles of Chiron and 2013~XZ8's PSFs were constructed using the photutils package. Separate radial profiles were computed across two wavelength ranges of interest: (1) the continuum region between 2.1 and 2.5~\um, which primarily probes for refractories such as dust grains, and (2) the region spanning the two \cotwo\ ice overtone absorption bands between 2.65 and 2.80~\um, to probe for solid \cotwo\ grains within the coma. For each wavelength range, the IFU slices corresponding to that region were collapsed along the wavelength axis to produce a median frame, from which the radial profile was derived. The resultant normalized profiles are plotted in Fig.~\ref{fig:sm2}. In both wavelength regions, the PSFs of Chiron and 2013~XZ8 display statistically indistinguishable radial profiles. No flux excess is detected on Chiron at any radial separation.

\subsection{Coma modeling}\label{subsec:comamod}

The Planetary Spectrum Generator (PSG) \citep{villanueva2018,villanueva2022} was used to produce models of the gas emission and derive estimates of the production rates of \methane\ and \cotwo. The position of Chiron at the time of the observations was imported via the JPL Horizons query function embedded within the browser-based application. Given that the irradiance spectrum of Chiron has been shifted into the target's stationary frame, the relative velocity value was manually adjusted to zero. The predefined instrument template for NIRSpec's high spectral resolution gratings was selected, and the Gaussian smoothing kernel was utilized for downsampling the output model spectra to the spectral resolution of the data. The telescope beam (i.e., the area over which the coma model is integrated) was set to match the $0.5'' \times 0.5''$ square extraction aperture. Surface continuum modeling was manually turned off in order to examine the gas emission exclusively. All other settings were left at their default values.

Coma modeling was initialized by selecting the predefined ``Expanding coma'' template. PSG employs the Cometary Emission Model (CEM), which accounts for the time-dependent expansion and evolution of gas molecules through the coma, relevant photodissociation lifetimes, collisionally induced absorption, and fluorescence pumping rates. The underlying theoretical formalism and implementation of the CEM are described in detail in the PSG Handbook \citep{villanueva2022}. For a given coma rotational temperature and aperture size, the production rate is degenerate with the coma expansion velocity. To remove this degeneracy, the expansion velocity at Chiron's heliocentric distance ($r = 18.8$~AU) was fixed to the value calculated from the empirical relation of Delsemme \citep{delsemme1982}: $580~\mathrm{m}~\mathrm{s}^{-1}\times r^{-0.5} = 130~\mathrm{m}~\mathrm{s}^{-1}$. The line lists for the gas-phase coma molecules were drawn from the GSFC Fluorescence Database \citep{villanueva2011}.

\begin{figure*}[ht]%
\centering
\includegraphics[width=\textwidth]{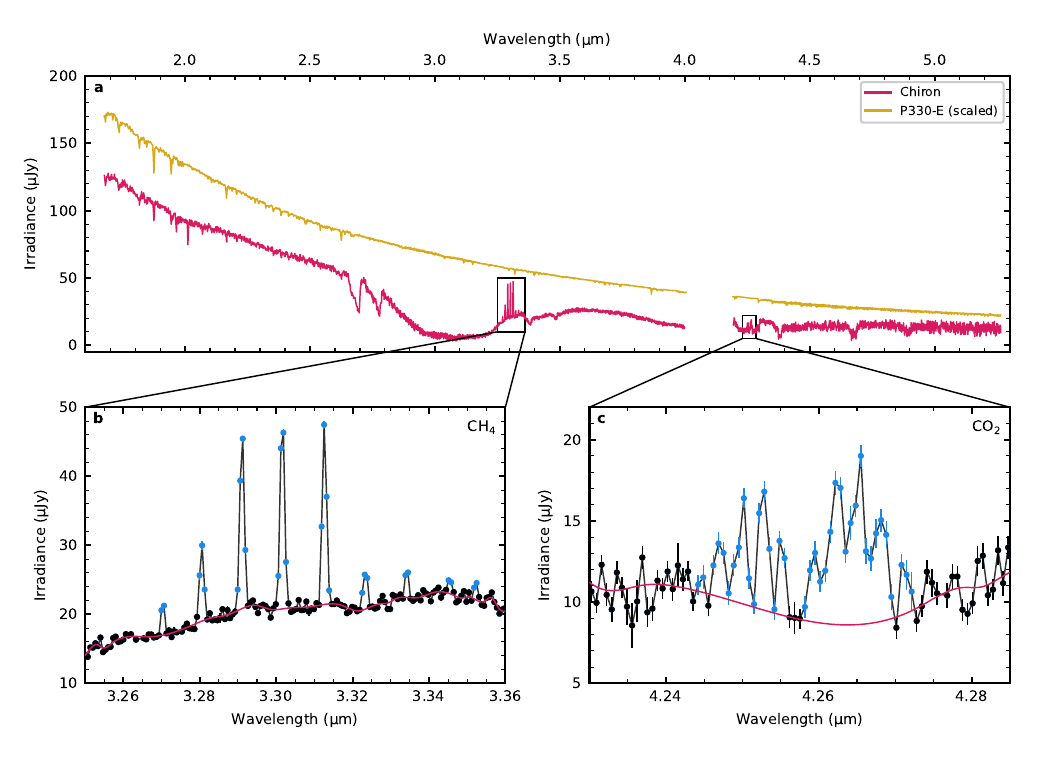}
\caption{\textbf{Extracted irradiance spectra and continuum removal.} \textbf{a} Irradiance spectra of Chiron (magenta) and the solar analog P330-E (yellow, scaled) extracted from the fully calibrated NIRSpec IFU data cubes using a $5\times5$~pixel box centered on the centroid pixel. \textbf{b} Zoomed-in view of the \methane\ fluorescence region, with the data points associated with the fluorescence marked in cyan and the fitted continuum spline model overplotted in magenta. \textbf{c} Same as panel \textbf{b}, but for the \cotwo\ fluorescence region.}\label{fig:sm1}
\end{figure*}

To remove the continuum contribution in Chiron's irradiance spectrum and isolate the gas emission bands, forward models were computed using PSG to identify the wavelengths at which gas fluorescence is present. For \methane, the emission peaks are fully resolved, and the corresponding points are marked in blue in Fig.~\ref{fig:sm1}b. After masking the points within the emission peaks, the basis spline interpolation routine \texttt{scipy.interpolate.splrep} was used to model the local continuum with a cubic spline function. The weights on the individual points were fixed to the inverse of the flux uncertainties. The convergence condition was set to $\chi^2 = 2N$, where $\chi^2$ is the error-weighted chi-squared metric, and $N$ is the number of points in the trimmed spectrum. This threshold was chosen to balance the closeness of the continuum fit with the smoothness of the spline interpolation. The continuum fit is shown by the red curve in Fig.~\ref{fig:sm1}b. The spectrum was restricted to the range 3.25--3.36~\um\ before being uploaded to PSG.

The lower signal-to-noise ratio of Chiron's spectrum in the vicinity of the \cotwo\ emission bands made continuum removal more challenging. As indicated in Fig.~\ref{fig:sm1}c, all points spanning the emission band region were masked, with the exception of the three points separating the P- and R-branches and the single points surrounding the two weakest emission bands. These retained points allowed the spline fit to more accurately model the local continuum shape. While alterations to the number of retained continuum points did not significantly affect the measured \cotwo\ production rate, this prescription was critical in the pixel-by-pixel coma mapping analysis, described below. The spectrum was trimmed to the wavelength range 4.23--4.29~\um, and the same spline fitting parameter settings were used as in the case of \methane.

\methane\ and \cotwo\ coma models were fit to the continuum-removed spectrum segments using the optimal estimation procedure in PSG's retrieval module; the coma temperature and production rate were the only free parameters. The $1\sigma$ uncertainties on the parameter values were estimated using the Jacobian calculated near the best-fit solution.

For CO, no distinct emission bands are discernible above the level of noise in the spectrum. The continuum and CO ice absorption band profile were modeled and removed using the same spline fitting routine, and the resultant 4.60--4.75~\um\ spectrum was fit with a CO coma model to derive the upper limit on the CO production rate.

\subsection{Coma mapping}\label{subsec:comamap}

The dither-combined G395H data cube was cleaned using a similar outlier cleaning procedure to the method used on the individual dither data cubes. For pixels outside of the $11\times11$ pixel box centered on the centroid, $5\sigma$ outliers across each slice were iteratively masked. For the remaining central pixels, points in the individual pixel spectra that differed by $>10\sigma$ from the cubic spline were flagged. Next, all individual pixel spectra in the vicinity of the \methane\ and \cotwo\ emission bands were carefully inspected, with lower-level outliers manually masked to produce the final cleaned data cube.

The map of \methane\ emission band area was computed pixel-by-pixel by fitting the local continuum using the same spline interpolation technique described in the previous section, subtracting the spline function from the pixel spectrum, and numerically integrating the flux across the 3.265--3.340~\um\ region using Simpson's rule. The significantly lower precision and larger scatter in the pixel spectra near the \cotwo\ emission bands necessitated a simplified approach, as the spline interpolation was less robust when modeling the local continuum. Instead, the designated continuum points between 4.24 and 4.28~\um\ were fit with a linear function; here, the anchor points located among the \cotwo\ band peaks were critical for ensuring reliable continuum fits.

\subsection{Surface composition modeling
}\label{subsec:reflmod}

Chiron's surface composition was investigated using a radiative transfer model based on the Hapke formalism \citep{hapke1993, hapke2002, hapke2012}. The bidirectional reflectance, expressed as the radiance factor ($I/F$), was computed by accounting for multiple scattering within a particulate medium, the shadow-hiding opposition effect, and macroscopic surface roughness. The numerical implementation was built upon previously developed modeling frameworks \citep{protopapa2017, protopapa2020, protopapa2025}. A range of surface configurations was explored, including areal mixtures, intimate mixtures, and vertically-stratified surfaces. 

The main free parameters in the model were the grain size ($D$) and relative abundance of each compound. The latter was expressed as either fractional area ($F$) or fractional volume ($V$) for areal or intimate mixtures, respectively. Parameter estimation followed the two-stage optimization strategy described in \cite{protopapa2025}, combining an initial Levenberg--Marquardt (LM) least-squares minimization process with a subsequent Markov Chain Monte Carlo (MCMC) exploration of the parameter space. The LM solution provided the starting point for the MCMC analysis, from which posterior probability distributions and credible intervals were derived. A final LM minimization, initialized with the median posterior values, was performed to refine the residuals while preserving the statistically informed parameter estimates. 

Only the single-scattering albedo, $w$, was assumed to vary with wavelength. The remaining Hapke photometric parameters were fixed to the values derived by \cite{verbiscer2022} for 2002~MS4: a shadow-hiding opposition surge amplitude and width of $B_0 = 1.0$ and $h = 0.24$, respectively, and a macroscopic roughness angle of $\theta = 20^\circ$. The double-lobed Henyey--Greenstein phase function was adopted, with a relative forward-scattering amplitude of $c=0.65$. The asymmetry parameter, $\xi$, which characterizes the angular distribution of scattered light, was treated as a free parameter. 

For configurations including aggregates, additional free parameters were introduced to describe the relative volume fractions of the individual constituents within the aggregate. The effective complex refractive index as a function of wavelength was computed using effective medium theory, specifically the Bruggeman mixing formalism \cite{bohren1983}. This approach preserves the spectral characteristics of the individual endmembers under the assumption that the components remain physically distinct at the molecular scale. The resulting effective optical constants were then used as inputs to the Hapke radiative transfer model. 

The best-fit model presented in Section~\ref{subsec:surface} was obtained using crystalline \water\ ice optical constants at 60~K \cite{mastrapa2009} mixed in aggregate form with \cotwo\ ice optical constants \cite{hansen1997}. The remaining components consisted of amorphous \water\ ice at 60~K \cite{mastrapa2009} and Titan tholin \cite{khare1984}.

\begin{figure*}[ht]%
\centering
\includegraphics[width=\textwidth]{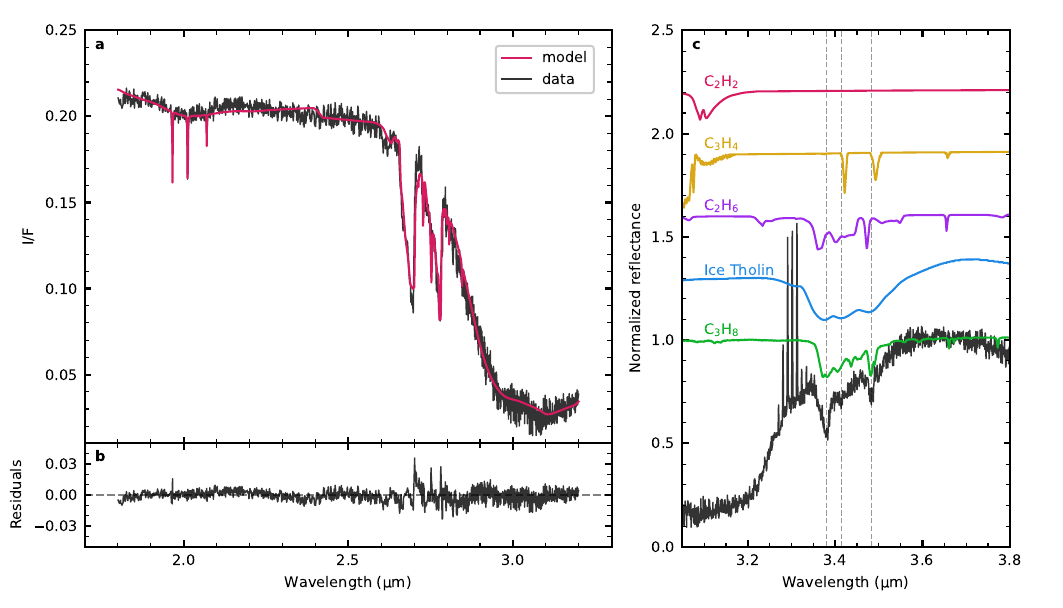}
\caption{\textbf{Spectral characterization of Chiron’s surface composition.} \textbf{a} Comparison of the best-fit spectral model of the surface composition (red) with the measured spectrum (black). The surface is best described by an intimate mixture of water ice in both amorphous and crystalline phases, carbon dioxide ice, and tholin-like materials. \textbf{b} Residuals between the model and the data show an excellent fit, with a root mean square deviation that is just 1\% higher than the average scatter in the data. \textbf{c} Zoomed-in view of the complex organic absorption feature spanning 3.3--3.6~\um, with the individual component bands indicated by dashed lines. Comparison between the measured spectrum and synthetic spectra of several candidate species, convolved to the spectral resolution of the data, points to C$_{3}$H$_{8}$ and/or ice tholins produced by the irradiation of \water\ ice and C$_{2}$H$_{6}$ as the most plausible candidates for the observed organic feature. } \label{fig:2}
\end{figure*}

To identify the species responsible for the complex organic absorption feature detected between 3.3 and 3.6~\um, the observed spectrum was compared with synthetic spectra of various candidate compounds (Fig.~\ref{fig:2}c). Specifically, synthetic hydrocarbon models were generated using the same framework described above, assuming intimate mixtures of amorphous carbon and the hydrocarbon species under investigation. The synthetic spectra were computed assuming an 80\% volume fraction of amorphous carbon and a 20\% volume fraction of the hydrocarbon, with a particle diameter of 5~\um. Optical constants for C$_3$H$_8$ and C$_3$H$_4$ were taken from \cite{hudson2021}, while data for C$_2$H$_2$ were retrieved from \citep{hudson2014}; C$_2$H$_6$ optical constants were obtained from R.~Mastrapa (private communication; see also \citep{protopapa2025}). For ice tholins, the model was constructed assuming a pure component using the optical constants of \cite{khare1993}.

\bibliography{sn-bibliography}

@ARTICLE{levison1997,
       author = {{Levison}, Harold F. and {Duncan}, Martin J.},
        title = "{From the Kuiper Belt to Jupiter-Family Comets: The Spatial Distribution of Ecliptic Comets}",
      journal = {\icarus},
         year = 1997,
        month = may,
       volume = {127},
       number = {1},
        pages = {13-32},
          doi = {10.1006/icar.1996.5637},
       adsurl = {https://ui.adsabs.harvard.edu/abs/1997Icar..127...13L}
}

@ARTICLE{tiscareno2003,
       author = {{Tiscareno}, Matthew S. and {Malhotra}, Renu},
        title = "{The Dynamics of Known Centaurs}",
      journal = {\aj},
         year = 2003,
        month = dec,
       volume = {126},
       number = {6},
        pages = {3122-3131},
          doi = {10.1086/379554},
archivePrefix = {arXiv},
       eprint = {astro-ph/0211076},
 primaryClass = {astro-ph},
       adsurl = {https://ui.adsabs.harvard.edu/abs/2003AJ....126.3122T}
}

@ARTICLE{nesvorny2017,
       author = {{Nesvorn{\'y}}, David and {Vokrouhlick{\'y}}, David and {Dones}, Luke and {Levison}, Harold F. and {Kaib}, Nathan and {Morbidelli}, Alessandro},
        title = "{Origin and Evolution of Short-period Comets}",
      journal = {\apj},
         year = 2017,
        month = aug,
       volume = {845},
       number = {1},
          eid = {27},
        pages = {27},
          doi = {10.3847/1538-4357/aa7cf6},
archivePrefix = {arXiv},
       eprint = {1706.07447},
 primaryClass = {astro-ph.EP},
       adsurl = {https://ui.adsabs.harvard.edu/abs/2017ApJ...845...27N}
}

@ARTICLE{pereira2025,
       author = {{Pereira}, C.~L. and {Braga-Ribas}, F. and {Sicardy}, B. and {Leiva}, R. and {Assafin}, M. and {Morgado}, B.~E. and {Ortiz}, J.~L. and {Santos-Sanz}, P. and {Camargo}, J.~I.~B. and {Margoti}, G. and {Kilic}, Y. and {Benedetti-Rossi}, G. and {Vieira-Martins}, R. and {Pinheiro}, T.~F.~L.~L. and {Sfair}, R. and {Rommel}, F.~L. and {Gomes-J{\'u}nior}, A.~R. and {Boufleur}, R.~C. and {Duffard}, R. and {Desmars}, J. and {Souami}, D. and {Morales}, N. and {Arrese}, F. and {Barkaoui}, K. and {Burdanov}, A. and {Colazo}, C.~A. and {Domingues}, C.~A. and {Dutra}, H. and {Gargalhone}, R.~C. and {Jacques}, C. and {Jablonski}, F. and {Liberato}, L. and {Melia}, R. and {Oliveira}, J.~C. and {Sardi{\~n}a}, M. and {Spagnotto}, J. and {Speranza}, T. and {Wilberger}, A. and {Zorzan}, M.~A. and {Brito}, L.~S. and {Cavalcante}, J.~P. and {Costa}, T.~Q. and {Emilio}, M. and {Garcia-Migani}, E. and {Gillon}, M. and {Gradovski}, E. and {Jehin}, E. and {Lattari}, V. and {Malacarne}, M. and {Mammana}, L.~A. and {Melita}, M. and {Melo}, W. and {Ortiz}, A.~J. and {Quitral-Manosalva}, P. and {Ramon}, G. and {Rodrigues}, I. and {Vanzi}, L.},
        title = "{The Rings of (2060) Chiron: Evidence of an Evolving System}",
      journal = {\apjl},
         year = 2025,
        month = oct,
       volume = {992},
       number = {2},
          eid = {L19},
        pages = {L19},
          doi = {10.3847/2041-8213/ae0b6d},
archivePrefix = {arXiv},
       eprint = {2510.12388},
 primaryClass = {astro-ph.EP},
       adsurl = {https://ui.adsabs.harvard.edu/abs/2025ApJ...992L..19P}
}

@ARTICLE{pinillaalonso2024,
       author = {{Pinilla-Alonso}, N. and {Licandro}, J. and {Brunetto}, R. and {Henault}, E. and {Schambeau}, C. and {Guilbert-Lepoutre}, A. and {Stansberry}, J. and {Wong}, I. and {Lunine}, J.~I. and {Holler}, B.~J. and {Emery}, J. and {Protopapa}, S. and {Cook}, J. and {Hammel}, H.~B. and {Villanueva}, G.~L. and {Milam}, S.~N. and {Cruikshank}, D. and {de Souza-Feliciano}, A.~C.},
        title = "{Unveiling the ice and gas nature of active centaur (2060) Chiron using the James Webb Space Telescope}",
      journal = {\aap},
         year = 2024,
        month = dec,
       volume = {692},
          eid = {L11},
        pages = {L11},
          doi = {10.1051/0004-6361/202450124},
archivePrefix = {arXiv},
       eprint = {2407.07761},
 primaryClass = {astro-ph.EP},
       adsurl = {https://ui.adsabs.harvard.edu/abs/2024A&A...692L..11P}
}

@ARTICLE{dobson2024,
       author = {{Dobson}, Matthew M. and {Schwamb}, Megan E. and {Fitzsimmons}, Alan and {Schambeau}, Charles and {Beck}, Aren and {Denneau}, Larry and {Erasmus}, Nicolas and {Heinze}, A.~N. and {Shingles}, Luke J. and {Siverd}, Robert J. and {Smith}, Ken W. and {Tonry}, John L. and {Weiland}, Henry and {Young}, David. R. and {Kelley}, Michael S.~P. and {Lister}, Tim and {Bernardinelli}, Pedro H. and {Ferrais}, Marin and {Jehin}, Emmanuel and {Fedorets}, Grigori and {Benecchi}, Susan D. and {Verbiscer}, Anne J. and {Murtagh}, Joseph and {Duffard}, Ren{\'e} and {Gomez}, Edward and {Chatelain}, Joey and {Greenstreet}, Sarah},
        title = "{The Discovery and Evolution of a Possible New Epoch of Cometary Activity by the Centaur (2060) Chiron}",
      journal = {\psj},
         year = 2024,
        month = jul,
       volume = {5},
       number = {7},
          eid = {165},
        pages = {165},
          doi = {10.3847/PSJ/ad543c},
archivePrefix = {arXiv},
       eprint = {2407.14410},
 primaryClass = {astro-ph.EP},
       adsurl = {https://ui.adsabs.harvard.edu/abs/2024PSJ.....5..165D}
}

@ARTICLE{sickafoose2023,
       author = {{Sickafoose}, Amanda A. and {Levine}, Stephen E. and {Bosh}, Amanda S. and {Person}, Michael J. and {Zuluaga}, Carlos A. and {Knieling}, Bastian and {Lewis}, Mark C. and {Schindler}, Karsten},
        title = "{Material around the Centaur (2060) Chiron from the 2018 November 28 UT Stellar Occultation}",
      journal = {\psj},
         year = 2023,
        month = nov,
       volume = {4},
       number = {11},
          eid = {221},
        pages = {221},
          doi = {10.3847/PSJ/ad0632},
archivePrefix = {arXiv},
       eprint = {2310.16205},
 primaryClass = {astro-ph.EP},
       adsurl = {https://ui.adsabs.harvard.edu/abs/2023PSJ.....4..221S}
}

@ARTICLE{ortiz2023,
       author = {{Ortiz}, J.~L. and {Pereira}, C.~L. and {Sicardy}, B. and {Braga-Ribas}, F. and {Takey}, A. and {Fouad}, A.~M. and {Shaker}, A.~A. and {Kaspi}, S. and {Brosch}, N. and {Kretlow}, M. and {Leiva}, R. and {Desmars}, J. and {Morgado}, B.~E. and {Morales}, N. and {Vara-Lubiano}, M. and {Santos-Sanz}, P. and {Fern{\'a}ndez-Valenzuela}, E. and {Souami}, D. and {Duffard}, R. and {Rommel}, F.~L. and {Kilic}, Y. and {Erece}, O. and {Koseoglu}, D. and {Ege}, E. and {Morales}, R. and {Alvarez-Candal}, A. and {Rizos}, J.~L. and {G{\'o}mez-Lim{\'o}n}, J.~M. and {Assafin}, M. and {Vieira-Martins}, R. and {Gomes-J{\'u}nior}, A.~R. and {Camargo}, J.~I.~B. and {Lecacheux}, J.},
        title = "{Changing material around (2060) Chiron revealed by an occultation on December 15, 2022}",
      journal = {\aap},
         year = 2023,
        month = aug,
       volume = {676},
          eid = {L12},
        pages = {L12},
          doi = {10.1051/0004-6361/202347025},
archivePrefix = {arXiv},
       eprint = {2308.03458},
 primaryClass = {astro-ph.EP},
       adsurl = {https://ui.adsabs.harvard.edu/abs/2023A&A...676L..12O}
}

@ARTICLE{hartmann1990,
       author = {{Hartmann}, W.~K. and {Tholen}, D.~J. and {Meech}, K.~J. and {Cruikshank}, D.~P.},
        title = "{2060 Chiron: Colorimetry and cometary behavior}",
      journal = {\icarus},
         year = 1990,
        month = jan,
       volume = {83},
       number = {1},
        pages = {1-15},
          doi = {10.1016/0019-1035(90)90002-Q},
       adsurl = {https://ui.adsabs.harvard.edu/abs/1990Icar...83....1H}
}

@ARTICLE{gehrels1977,
       author = {{Gehrels}, T. and {Vesely}, C.~D. and {Sather}, R. and {Green}, R. and {Kowal}, C.~T. and {Marsden}, B.~G.},
        title = "{Slow-Moving Object Kowal}",
      journal = {IAU Circ.},
         year = 1977,
        month = nov,
       volume = {3130},
        pages = {1},
       adsurl = {https://ui.adsabs.harvard.edu/abs/1977IAUC.3130....1G}
}

@ARTICLE{bragaribas2023,
       author = {{Braga-Ribas}, F. and {Pereira}, C.~L. and {Sicardy}, B. and {Ortiz}, J.~L. and {Desmars}, J. and {Sickafoose}, A. and {Emilio}, M. and {Morgado}, B. and {Margoti}, G. and {Rommel}, F.~L. and {Camargo}, J.~I.~B. and {Assafin}, M. and {Vieira-Martins}, R. and {Gomes-J{\'u}nior}, A.~R. and {Santos-Sanz}, P. and {Morales}, N. and {Kretlow}, M. and {Lecacheux}, J. and {Colas}, F. and {Boninsegna}, R. and {Schreurs}, O. and {Dauvergne}, J.~L. and {Fernandez}, E. and {van Heerden}, H.~J. and {Gonz{\'a}lez}, H. and {Bihel}, D. and {Jankowsky}, F.},
        title = "{Constraints on (2060) Chiron's size, shape, and surrounding material from the November 2018 and September 2019 stellar occultations}",
      journal = {\aap},
         year = 2023,
        month = aug,
       volume = {676},
          eid = {A72},
        pages = {A72},
          doi = {10.1051/0004-6361/202346749},
archivePrefix = {arXiv},
       eprint = {2308.10042},
 primaryClass = {astro-ph.EP},
       adsurl = {https://ui.adsabs.harvard.edu/abs/2023A&A...676A..72B}
}

@ARTICLE{sickafoose2020,
       author = {{Sickafoose}, A.~A. and {Bosh}, A.~S. and {Emery}, J.~P. and {Person}, M.~J. and {Zuluaga}, C.~A. and {Womack}, M. and {Ruprecht}, J.~D. and {Bianco}, F.~B. and {Zangari}, A.~M.},
        title = "{Characterization of material around the centaur (2060) Chiron from a visible and near-infrared stellar occultation in 2011}",
      journal = {\mnras},
         year = 2020,
        month = jan,
       volume = {491},
       number = {3},
        pages = {3643-3654},
          doi = {10.1093/mnras/stz3079},
archivePrefix = {arXiv},
       eprint = {1910.05029},
 primaryClass = {astro-ph.EP},
       adsurl = {https://ui.adsabs.harvard.edu/abs/2020MNRAS.491.3643S}
}

@ARTICLE{lellouch2017,
       author = {{Lellouch}, E. and {Moreno}, R. and {M{\"u}ller}, T. and {Fornasier}, S. and {Santos-Sanz}, P. and {Moullet}, A. and {Gurwell}, M. and {Stansberry}, J. and {Leiva}, R. and {Sicardy}, B. and {Butler}, B. and {Boissier}, J.},
        title = "{The thermal emission of Centaurs and trans-Neptunian objects at millimeter wavelengths from ALMA observations}",
      journal = {\aap},
         year = 2017,
        month = dec,
       volume = {608},
          eid = {A45},
        pages = {A45},
          doi = {10.1051/0004-6361/201731676},
archivePrefix = {arXiv},
       eprint = {1709.06747},
 primaryClass = {astro-ph.EP},
       adsurl = {https://ui.adsabs.harvard.edu/abs/2017A&A...608A..45L}
}

@ARTICLE{womack2017,
       author = {{Womack}, M. and {Sarid}, G. and {Wierzchos}, K.},
        title = "{CO and Other Volatiles in Distantly Active Comets}",
      journal = {\pasp},
         year = 2017,
        month = mar,
       volume = {129},
       number = {973},
        pages = {031001},
          doi = {10.1088/1538-3873/129/973/031001},
archivePrefix = {arXiv},
       eprint = {1611.00051},
 primaryClass = {astro-ph.EP},
       adsurl = {https://ui.adsabs.harvard.edu/abs/2017PASP..129c1001W}
}

@ARTICLE{ruprecht2015,
       author = {{Ruprecht}, Jessica D. and {Bosh}, Amanda S. and {Person}, Michael J. and {Bianco}, Federica B. and {Fulton}, Benjamin J. and {Gulbis}, Amanda A.~S. and {Bus}, Schelte J. and {Zangari}, Amanda M.},
        title = "{29 November 2011 stellar occultation by 2060 Chiron: Symmetric jet-like features}",
      journal = {\icarus},
         year = 2015,
        month = may,
       volume = {252},
        pages = {271-276},
          doi = {10.1016/j.icarus.2015.01.015},
       adsurl = {https://ui.adsabs.harvard.edu/abs/2015Icar..252..271R}
}

@ARTICLE{ortiz2015,
       author = {{Ortiz}, J.~L. and {Duffard}, R. and {Pinilla-Alonso}, N. and {Alvarez-Candal}, A. and {Santos-Sanz}, P. and {Morales}, N. and {Fern{\'a}ndez-Valenzuela}, E. and {Licandro}, J. and {Campo Bagatin}, A. and {Thirouin}, A.},
        title = "{Possible ring material around centaur (2060) Chiron}",
      journal = {\aap},
         year = 2015,
        month = apr,
       volume = {576},
          eid = {A18},
        pages = {A18},
          doi = {10.1051/0004-6361/201424461},
archivePrefix = {arXiv},
       eprint = {1501.05911},
 primaryClass = {astro-ph.EP},
       adsurl = {https://ui.adsabs.harvard.edu/abs/2015A&A...576A..18O}
}

@ARTICLE{luu2000,
       author = {{Luu}, Jane X. and {Jewitt}, David C. and {Trujillo}, Chad},
        title = "{Water Ice in 2060 Chiron and Its Implications for Centaurs and Kuiper Belt Objects}",
      journal = {\apjl},
         year = 2000,
        month = mar,
       volume = {531},
       number = {2},
        pages = {L151-L154},
          doi = {10.1086/312536},
archivePrefix = {arXiv},
       eprint = {astro-ph/0002094},
 primaryClass = {astro-ph},
       adsurl = {https://ui.adsabs.harvard.edu/abs/2000ApJ...531L.151L}
}

@ARTICLE{womack1999,
       author = {{Womack}, Maria and {Stern}, S. Alan},
        title = "{The Detection of Carbon Monoxide Gas Emission in (2060) Chiron}",
      journal = {Solar System Research},
         year = 1999,
        month = jan,
       volume = {33},
        pages = {187},
       adsurl = {https://ui.adsabs.harvard.edu/abs/1999SoSyR..33..187W}
}

@ARTICLE{bus1996,
       author = {{Bus}, Schelte J. and {Buie}, Marc W. and {Schleicher}, David G. and {Hubbard}, William B. and {Marcialis}, Robert L. and {Hill}, Richard and {Wasserman}, Lawrence H. and {Spencer}, John R. and {Millis}, Robert L. and {Franz}, Otto G. and {Bosh}, Amanda S. and {Dunham}, Edward W. and {Ford}, Charles H. and {Young}, James W. and {Elliott}, J.~L. and {Meserole}, Richard and {Olkin}, Catherine B. and {McDonald}, Stephen W. and {Foust}, Jeffrey A. and {Sopata}, Lisa M. and {Bandyopadhyay}, Reba M.},
        title = "{Stellar Occultation by 2060 Chiron}",
      journal = {\icarus},
         year = 1996,
        month = oct,
       volume = {123},
       number = {2},
        pages = {478-490},
          doi = {10.1006/icar.1996.0173},
       adsurl = {https://ui.adsabs.harvard.edu/abs/1996Icar..123..478B}
}

@ARTICLE{prialnik1995,
       author = {{Prialnik}, Dina and {Brosch}, Noah and {Ianovici}, Drora},
        title = "{Modelling the activity of 2060 Chiron}",
      journal = {\mnras},
         year = 1995,
        month = oct,
       volume = {276},
       number = {4},
        pages = {1148-1154},
          doi = {10.1093/mnras/276.4.1148},
       adsurl = {https://ui.adsabs.harvard.edu/abs/1995MNRAS.276.1148P}
}

@ARTICLE{elliot1995,
       author = {{Elliot}, J.~L. and {Olkin}, C.~B. and {Dunham}, E.~W. and {Ford}, C.~H. and {Gilmore}, D.~K. and {Kurtz}, D. and {Lazzaro}, D. and {Rank}, D.~M. and {Temi}, P. and {Bandyopadhyay}, R.~M. and {Barroso}, J. and {Barucci}, A. and {Bosh}, A.~S. and {Buie}, M.~W. and {Bus}, S.~J. and {Dahn}, C.~C. and {Foryta}, D.~W. and {Hubbard}, W.~B. and {Lopes}, D.~F. and {Marcialis}, R.~L. and {McDonald}, S.~W. and {Millis}, R.~L. and {Reitsema}, H. and {Schleicher}, D.~G. and {Sicardy}, B. and {Stone}, R.~P.~S. and {Wasserman}, L.~H.},
        title = "{Jet-like features near the nucleus of Chiron}",
      journal = {\nat},
         year = 1995,
        month = jan,
       volume = {373},
       number = {6509},
        pages = {46-49},
          doi = {10.1038/373046a0},
       adsurl = {https://ui.adsabs.harvard.edu/abs/1995Natur.373...46E}
}

@ARTICLE{bus1991,
       author = {{Bus}, S.~J. and {A'Hearn}, M.~F. and {Schleicher}, D.~G. and {Bowell}, E.},
        title = "{Detection of CN Emission from (2060) Chiron}",
      journal = {Science},
         year = 1991,
        month = feb,
       volume = {251},
       number = {4995},
        pages = {774-777},
          doi = {10.1126/science.251.4995.774},
       adsurl = {https://ui.adsabs.harvard.edu/abs/1991Sci...251..774B}
}

@ARTICLE{meech1990,
       author = {{Meech}, Karen J. and {Belton}, Michael J.~S.},
        title = "{The Atmosphere of 2060 Chiron}",
      journal = {\aj},
         year = 1990,
        month = oct,
       volume = {100},
        pages = {1323},
          doi = {10.1086/115600},
       adsurl = {https://ui.adsabs.harvard.edu/abs/1990AJ....100.1323M}
}

@ARTICLE{luu1990,
       author = {{Luu}, Jane X. and {Jewitt}, David C.},
        title = "{Cometary Activity in 2060 Chiron}",
      journal = {\aj},
         year = 1990,
        month = sep,
       volume = {100},
        pages = {913},
          doi = {10.1086/115571},
       adsurl = {https://ui.adsabs.harvard.edu/abs/1990AJ....100..913L}
}

@ARTICLE{romanishin2018,
       author = {{Romanishin}, W. and {Tegler}, S.~C.},
        title = "{Albedos of Centaurs, Jovian Trojans, and Hildas}",
      journal = {\aj},
         year = 2018,
        month = jul,
       volume = {156},
       number = {1},
          eid = {19},
        pages = {19},
          doi = {10.3847/1538-3881/aac210},
       adsurl = {https://ui.adsabs.harvard.edu/abs/2018AJ....156...19R}
}

@ARTICLE{campins1994,
       author = {{Campins}, H. and {Telesco}, C.~M. and {Osip}, D.~J. and {Rieke}, G.~H. and {Rieke}, M.~J. and {Schulz}, B.},
        title = "{The Color Temperature of (2060) Chiron: A Warm and Small Nucleus}",
      journal = {\aj},
         year = 1994,
        month = dec,
       volume = {108},
        pages = {2318},
          doi = {10.1086/117244},
       adsurl = {https://ui.adsabs.harvard.edu/abs/1994AJ....108.2318C}
}

@ARTICLE{foster1999,
       author = {{Foster}, Michael J. and {Green}, Simon F. and {McBride}, Neil and {Davies}, John K.},
        title = "{NOTE: Detection of Water Ice on 2060 Chiron}",
      journal = {\icarus},
         year = 1999,
        month = oct,
       volume = {141},
       number = {2},
        pages = {408-410},
          doi = {10.1006/icar.1999.6180},
       adsurl = {https://ui.adsabs.harvard.edu/abs/1999Icar..141..408F}
}

@ARTICLE{tholen1988,
       author = {{Tholen}, D.~J. and {Hartmann}, W.~K. and {Cruikshank}, D.~P. and {Lilly}, S. and {Bowell}, E. and {Hewitt}, A.},
        title = "{(2060) Chiron}",
      journal = {IAU Circ.},
         year = 1988,
        month = feb,
       volume = {4554},
        pages = {2},
       adsurl = {https://ui.adsabs.harvard.edu/abs/1988IAUC.4554....2T}
}

@ARTICLE{cochran1991,
       author = {{Cochran}, A.~L. and {Cochran}, W.~D.},
        title = "{The first detection of CN and the distribution of CO $^{+}$ gas in the coma of Comet P/Schwassmann-Wachman 1}",
      journal = {\icarus},
         year = 1991,
        month = mar,
       volume = {90},
       number = {1},
        pages = {172-175},
          doi = {10.1016/0019-1035(91)90077-7},
       adsurl = {https://ui.adsabs.harvard.edu/abs/1991Icar...90..172C}
}

@BOOK{villanueva2022,
       author = {{Villanueva}, Geronimo Luis and {Liuzzi}, Giuliano and {Faggi}, Sara and {Protopapa}, Silvia and {Kofman}, Vincent and {Fauchez}, Thomas and {Stone}, Shane Wesley and {Mandell}, Avi Max},
        title = "{Fundamentals of the Planetary Spectrum Generator}",
         year = 2022,
       adsurl = {https://ui.adsabs.harvard.edu/abs/2022fpsg.book.....V}
}

@ARTICLE{quirico1997,
       author = {{Quirico}, Eric and {Schmitt}, Bernard},
        title = "{Near-Infrared Spectroscopy of Simple Hydrocarbons and Carbon Oxides Diluted in Solid N$_{2}$ and as Pure Ices: Implications for Triton and Pluto}",
      journal = {\icarus},
         year = 1997,
        month = jun,
       volume = {127},
       number = {2},
        pages = {354-378},
          doi = {10.1006/icar.1996.5663},
       adsurl = {https://ui.adsabs.harvard.edu/abs/1997Icar..127..354Q}
}

@INPROCEEDINGS{wong2025,
       author = {{Wong}, I. and {Holler}, B.~J. and {Protopapa}, S. and {Lellouch}, E. and {Hines}, D.~C. and {Brunetto}, R. and {Cook}, J.~C. and {Emery}, J.~P. and {Grundy}, W.~M. and {Lunine}, J.~I. and {Pinilla-Alonso}, N. and {Stansberry}, J.~A. and {Milam}, S.~N. and {Hammel}, H.~B.},
        title = "{The Complex Surface and Atmosphere of Triton as Revealed by JWST}",
    booktitle = {LPI Contributions},
         year = 2025,
        editor = {none},
       series = {LPI Contributions},
       volume = {3059},
        month = jul,
          eid = {7047},
        pages = {7047},
       adsurl = {https://ui.adsabs.harvard.edu/abs/2025LPICo3059.7047W}
}

@BOOK{hapke1993,
       author = {{Hapke}, Bruce},
        title = "{Theory of Reflectance and Emittance Spectroscopy}",
         year = 1993,
    publisher = {Cambridge University Press},
    address = {Cambridge, UK},
       adsurl = {https://ui.adsabs.harvard.edu/abs/1993tres.book.....H}
}

@BOOK{hapke2012,
       author = {{Hapke}, Bruce},
        title = "{Theory of Reflectance and Emittance Spectroscopy}",
         year = 2012,
          doi = {10.1017/CBO9781139025683},
    publisher = {Cambridge University Press},
    address = {Cambridge, UK},
       adsurl = {https://ui.adsabs.harvard.edu/abs/2012tres.book.....H}
}

@ARTICLE{hapke2002,
       author = {{Hapke}, Bruce},
        title = "{Bidirectional Reflectance Spectroscopy. 5. The Coherent Backscatter Opposition Effect and Anisotropic Scattering}",
      journal = {\icarus},
         year = 2002,
        month = jun,
       volume = {157},
       number = {2},
        pages = {523-534},
          doi = {10.1006/icar.2002.6853},
       adsurl = {https://ui.adsabs.harvard.edu/abs/2002Icar..157..523H}
}

@ARTICLE{protopapa2017,
       author = {{Protopapa}, S. and {Grundy}, W.~M. and {Reuter}, D.~C. and {Hamilton}, D.~P. and {Dalle Ore}, C.~M. and {Cook}, J.~C. and {Cruikshank}, D.~P. and {Schmitt}, B. and {Philippe}, S. and {Quirico}, E. and {Binzel}, R.~P. and {Earle}, A.~M. and {Ennico}, K. and {Howett}, C.~J.~A. and {Lunsford}, A.~W. and {Olkin}, C.~B. and {Parker}, A. and {Singer}, K.~N. and {Stern}, A. and {Verbiscer}, A.~J. and {Weaver}, H.~A. and {Young}, L.~A. and {New Horizons Science Team}},
        title = "{Pluto's global surface composition through pixel-by-pixel Hapke modeling of New Horizons Ralph/LEISA data}",
      journal = {\icarus},
         year = 2017,
        month = may,
       volume = {287},
        pages = {218-228},
          doi = {10.1016/j.icarus.2016.11.028},
archivePrefix = {arXiv},
       eprint = {1604.08468},
 primaryClass = {astro-ph.EP},
       adsurl = {https://ui.adsabs.harvard.edu/abs/2017Icar..287..218P}
}

@ARTICLE{Mejia2015,
       author = {{Mej{\'\i}a}, C. and {Bender}, M. and {Severin}, D. and {Trautmann}, C. and {Boduch}, Ph. and {Bordalo}, V. and {Domaracka}, A. and {Lv}, X.~Y. and {Martinez}, R. and {Rothard}, H.},
        title = "{Radiolysis and sputtering of carbon dioxide ice induced by swift Ti, Ni, and Xe ions}",
      journal = {Nuclear Instruments and Methods in Physics Research B},
         year = 2015,
        month = dec,
       volume = {365},
        pages = {477-481},
          doi = {10.1016/j.nimb.2015.09.039},
       adsurl = {https://ui.adsabs.harvard.edu/abs/2015NIMPB.365..477M}
}

@ARTICLE{Brucato1997,
       author = {{Brucato}, J.~R. and {Palumbo}, M.~E. and {Strazzulla}, G.},
        title = "{Carbonic Acid by Ion Implantation in Water\_solarCarbon Dioxide Ice Mixtures}",
      journal = {\icarus},
         year = 1997,
        month = jan,
       volume = {125},
       number = {1},
        pages = {135-144},
          doi = {10.1006/icar.1996.5561},
       adsurl = {https://ui.adsabs.harvard.edu/abs/1997Icar..125..135B}
}

@ARTICLE{Quirico2023,
       author = {{Quirico}, Eric and {Bacmann}, Aurore and {Wolters}, C{\'e}dric and {Aug{\'e}}, Basile and {Flandinet}, Laur{\`e}ne and {Launois}, Thibault and {Cooper}, John F. and {Vuitton}, V{\'e}ronique and {Gautier}, Thomas and {Jovanovic}, Lora and {Boduch}, Philippe and {Rothard}, Hermann and {Desage}, L{\'e}opold and {Faure}, Alexandre and {Schmitt}, Bernard and {Poch}, Olivier and {Grundy}, William M. and {Protopapa}, Silvia and {Fornasier}, Sonia and {Cruikshank}, Dale P. and {Stern}, S. Alan and {New Horizons Team}},
        title = "{On a radiolytic origin of red organics at the surface of the Arrokoth Trans-Neptunian Object}",
      journal = {\icarus},
         year = 2023,
        month = apr,
       volume = {394},
          eid = {115396},
        pages = {115396},
          doi = {10.1016/j.icarus.2022.115396},
       adsurl = {https://ui.adsabs.harvard.edu/abs/2023Icar..39415396Q}
}

@ARTICLE{disanti2021,
       author = {{DiSanti}, Michael A. and {Bonev}, Boncho P. and {Dello Russo}, Neil and {McKay}, Adam J. and {Roth}, Nathan X. and {Saki}, Mohammad and {Gibb}, Erika L. and {Vervack}, Jr., Ronald J. and {Khan}, Younas and {Kawakita}, Hideyo},
        title = "{Volatile Composition and Outgassing in C/2018 Y1 (Iwamoto): Extending Limits for High-resolution Infrared Cometary Spectroscopy between 2.8 and 5.0 {\ensuremath{\mu}}m}",
      journal = {\psj},
         year = 2021,
        month = dec,
       volume = {2},
       number = {6},
          eid = {225},
        pages = {225},
          doi = {10.3847/PSJ/ac07ae},
       adsurl = {https://ui.adsabs.harvard.edu/abs/2021PSJ.....2..225D}
}

@ARTICLE{cabral2019,
       author = {{Cabral}, Nahuel and {Guilbert-Lepoutre}, Aur{\'e}lie and {Fraser}, Wesley C. and {Marsset}, Micha{\"e}l and {Volk}, Kathryn and {Petit}, Jean-Marc and {Rousselot}, Philippe and {Alexandersen}, Mike and {Bannister}, Michele T. and {Chen}, Ying-Tung and {Gladman}, Brett and {Gwyn}, Stephen D.~J. and {Kavelaars}, John J.},
        title = "{OSSOS. XI. No active centaurs in the Outer Solar System Origins Survey}",
      journal = {\aap},
         year = 2019,
        month = jan,
       volume = {621},
          eid = {A102},
        pages = {A102},
          doi = {10.1051/0004-6361/201834021},
archivePrefix = {arXiv},
       eprint = {1810.03648},
 primaryClass = {astro-ph.EP},
       adsurl = {https://ui.adsabs.harvard.edu/abs/2019A&A...621A.102C}
}

@ARTICLE{li2020,
       author = {{Li}, Jing and {Jewitt}, David and {Mutchler}, Max and {Agarwal}, Jessica and {Weaver}, Harold},
        title = "{Hubble Space Telescope Search for Activity in High-perihelion Objects}",
      journal = {\aj},
         year = 2020,
        month = may,
       volume = {159},
       number = {5},
          eid = {209},
        pages = {209},
          doi = {10.3847/1538-3881/ab7faf},
archivePrefix = {arXiv},
       eprint = {2003.06519},
 primaryClass = {astro-ph.EP},
       adsurl = {https://ui.adsabs.harvard.edu/abs/2020AJ....159..209L}
}

@ARTICLE{guilbertlepoutre2023,
       author = {{Guilbert-Lepoutre}, Aur{\'e}lie and {Gkotsinas}, Anastasios and {Raymond}, Sean N. and {Nesvorny}, David},
        title = "{The Gateway from Centaurs to Jupiter-family Comets: Thermal and Dynamical Evolution}",
      journal = {\apj},
         year = 2023,
        month = jan,
       volume = {942},
       number = {2},
          eid = {92},
        pages = {92},
          doi = {10.3847/1538-4357/acaa3a},
archivePrefix = {arXiv},
       eprint = {2212.06637},
 primaryClass = {astro-ph.EP},
       adsurl = {https://ui.adsabs.harvard.edu/abs/2023ApJ...942...92G}
}

@ARTICLE{roth2017,
       author = {{Roth}, Nathan X. and {Gibb}, Erika L. and {Bonev}, Boncho P. and {DiSanti}, Michael A. and {Mumma}, Michael J. and {Villanueva}, Geronimo L. and {Paganini}, Lucas},
        title = "{The Composition of Comet C/2012 K1 (PanSTARRS) and the Distribution of Primary Volatile Abundances among Comets}",
      journal = {\aj},
         year = 2017,
        month = apr,
       volume = {153},
       number = {4},
          eid = {168},
        pages = {168},
          doi = {10.3847/1538-3881/aa5d18},
       adsurl = {https://ui.adsabs.harvard.edu/abs/2017AJ....153..168R}
}

@ARTICLE{dellorusso2022,
       author = {{Dello Russo}, Neil and {Vervack}, Ronald J. and {Kawakita}, Hideyo and {Bonev}, Boncho P. and {DiSanti}, Michael A. and {Gibb}, Erika L. and {McKay}, Adam J. and {Cochran}, Anita L. and {Weaver}, Harold A. and {Biver}, Nicolas and {Crovisier}, Jacques and {Bockel{\'e}e-Morvan}, Dominique and {Kobayashi}, Hitomi and {Harris}, Walter M. and {Roth}, Nathan X. and {Saki}, Mohammad and {Khan}, Younas},
        title = "{Volatile Abundances, Extended Coma Sources, and Nucleus Ice Associations in Comet C/2014 Q2 (Lovejoy)}",
      journal = {\psj},
         year = 2022,
        month = jan,
       volume = {3},
       number = {1},
          eid = {6},
        pages = {6},
          doi = {10.3847/PSJ/ac323c},
       adsurl = {https://ui.adsabs.harvard.edu/abs/2022PSJ.....3....6D}
}

@ARTICLE{bonev2017,
       author = {{Bonev}, Boncho P. and {Villanueva}, Geronimo L. and {DiSanti}, Michael A. and {Boehnhardt}, Hermann and {Lippi}, Manuela and {Gibb}, Erika L. and {Paganini}, Lucas and {Mumma}, Michael J.},
        title = "{Beyond 3 au from the Sun: The Hypervolatiles CH$_{4}$, C$_{2}$H$_{6}$, and CO in the Distant Comet C/2006 W3 (Christensen)}",
      journal = {\aj},
         year = 2017,
        month = may,
       volume = {153},
       number = {5},
          eid = {241},
        pages = {241},
          doi = {10.3847/1538-3881/aa64dd},
       adsurl = {https://ui.adsabs.harvard.edu/abs/2017AJ....153..241B}
}

@ARTICLE{mckay2019,
       author = {{McKay}, Adam J. and {DiSanti}, Michael A. and {Kelley}, Michael S.~P. and {Knight}, Matthew M. and {Womack}, Maria and {Wierzchos}, Kacper and {Harrington Pinto}, Olga and {Bonev}, Boncho and {Villanueva}, Geronimo L. and {Dello Russo}, Neil and {Cochran}, Anita L. and {Biver}, Nicolas and {Bauer}, James and {Vervack}, Jr., Ronald J. and {Gibb}, Erika and {Roth}, Nathan and {Kawakita}, Hideyo},
        title = "{The Peculiar Volatile Composition of CO-dominated Comet C/2016 R2 (PanSTARRS)}",
      journal = {\aj},
         year = 2019,
        month = sep,
       volume = {158},
       number = {3},
          eid = {128},
        pages = {128},
          doi = {10.3847/1538-3881/ab32e4},
archivePrefix = {arXiv},
       eprint = {1907.07208},
 primaryClass = {astro-ph.EP},
       adsurl = {https://ui.adsabs.harvard.edu/abs/2019AJ....158..128M}
}

@ARTICLE{faggi2018,
       author = {{Faggi}, S. and {Villanueva}, G.~L. and {Mumma}, M.~J. and {Paganini}, L.},
        title = "{The Volatile Composition of Comet C/2017 E4 (Lovejoy) before its Disruption, as Revealed by High-resolution Infrared Spectroscopy with iSHELL at the NASA/IRTF}",
      journal = {\aj},
         year = 2018,
        month = aug,
       volume = {156},
       number = {2},
          eid = {68},
        pages = {68},
          doi = {10.3847/1538-3881/aace01},
       adsurl = {https://ui.adsabs.harvard.edu/abs/2018AJ....156...68F}
}

@ARTICLE{faggi2019,
       author = {{Faggi}, S. and {Mumma}, M.~J. and {Villanueva}, G.~L. and {Paganini}, L. and {Lippi}, M.},
        title = "{Quantifying the Evolution of Molecular Production Rates of Comet 21P/Giacobini-Zinner with iSHELL/NASA-IRTF}",
      journal = {\aj},
         year = 2019,
        month = dec,
       volume = {158},
       number = {6},
          eid = {254},
        pages = {254},
          doi = {10.3847/1538-3881/ab4f6e},
       adsurl = {https://ui.adsabs.harvard.edu/abs/2019AJ....158..254F}
}

@ARTICLE{faggi2021,
       author = {{Faggi}, Sara and {Lippi}, Manuela and {Camarca}, Maria and {Buzard}, Camillus F. and {Villanueva}, Geronimo L. and {Doppmann}, Gregory W. and {Blake}, Geoffrey A. and {Mumma}, Michael J.},
        title = "{The Extraordinary Passage of Comet C/2020 F3 NEOWISE: Evidence for Heterogeneous Chemical Inventory in Its Nucleus}",
      journal = {\aj},
         year = 2021,
        month = nov,
       volume = {162},
       number = {5},
          eid = {178},
        pages = {178},
          doi = {10.3847/1538-3881/ac179c},
       adsurl = {https://ui.adsabs.harvard.edu/abs/2021AJ....162..178F}
}

@ARTICLE{faggi2023,
       author = {{Faggi}, Sara and {Lippi}, Manuela and {Mumma}, Michael J. and {Villanueva}, Geronimo L.},
        title = "{Strongly Depleted Methanol and Hypervolatiles in Comet C/2021 A1 (Leonard): Signatures of Interstellar Chemistry?}",
      journal = {\psj},
         year = 2023,
        month = jan,
       volume = {4},
       number = {1},
          eid = {8},
        pages = {8},
          doi = {10.3847/PSJ/aca64c},
       adsurl = {https://ui.adsabs.harvard.edu/abs/2023PSJ.....4....8F}
}

@ARTICLE{disanti2017,
       author = {{DiSanti}, Michael A. and {Bonev}, Boncho P. and {Russo}, Neil Dello and {Vervack}, Jr., Ronald J. and {Gibb}, Erika L. and {Roth}, Nathan X. and {McKay}, Adam J. and {Kawakita}, Hideyo and {Feaga}, Lori M. and {Weaver}, Harold A.},
        title = "{Hypervolatiles in a Jupiter-family Comet: Observations of 45P/Honda-Mrkos-Pajdu{\v{s}}{\'a}kov{\'a} Using iSHELL at the NASA-IRTF}",
      journal = {\aj},
         year = 2017,
        month = dec,
       volume = {154},
       number = {6},
          eid = {246},
        pages = {246},
          doi = {10.3847/1538-3881/aa8639},
       adsurl = {https://ui.adsabs.harvard.edu/abs/2017AJ....154..246D}
}

@ARTICLE{disanti2018,
       author = {{DiSanti}, Michael A. and {Bonev}, Boncho P. and {Gibb}, Erika L. and {Roth}, Nathan X. and {Dello Russo}, Neil and {Vervack}, Jr., Ronald J.},
        title = "{Comet C/2013 V5 (Oukaimeden): Evidence for Depleted Organic Volatiles and Compositional Heterogeneity as Revealed through Infrared Spectroscopy}",
      journal = {\aj},
         year = 2018,
        month = dec,
       volume = {156},
       number = {6},
          eid = {258},
        pages = {258},
          doi = {10.3847/1538-3881/aade87},
       adsurl = {https://ui.adsabs.harvard.edu/abs/2018AJ....156..258D}
}

@ARTICLE{leroy2015,
       author = {{Le Roy}, L{\'e}na and {Altwegg}, Kathrin and {Balsiger}, Hans and {Berthelier}, Jean-Jacques and {Bieler}, Andre and {Briois}, Christelle and {Calmonte}, Ursina and {Combi}, Michael R. and {De Keyser}, Johan and {Dhooghe}, Frederik and {Fiethe}, Bj{\"o}rn and {Fuselier}, Stephen A. and {Gasc}, S{\'e}bastien and {Gombosi}, Tamas I. and {H{\"a}ssig}, Myrtha and {J{\"a}ckel}, Annette and {Rubin}, Martin and {Tzou}, Chia-Yu},
        title = "{Inventory of the volatiles on comet 67P/Churyumov-Gerasimenko from Rosetta/ROSINA}",
      journal = {\aap},
         year = 2015,
        month = nov,
       volume = {583},
          eid = {A1},
        pages = {A1},
          doi = {10.1051/0004-6361/201526450},
       adsurl = {https://ui.adsabs.harvard.edu/abs/2015A&A...583A...1L}
}

@ARTICLE{schaller2007,
       author = {{Schaller}, E.~L. and {Brown}, M.~E.},
        title = "{Volatile Loss and Retention on Kuiper Belt Objects}",
      journal = {\apjl},
         year = 2007,
        month = apr,
       volume = {659},
       number = {1},
        pages = {L61-L64},
          doi = {10.1086/516709},
       adsurl = {https://ui.adsabs.harvard.edu/abs/2007ApJ...659L..61S}
}

@ARTICLE{fray2009,
       author = {{Fray}, N. and {Schmitt}, B.},
        title = "{Sublimation of ices of astrophysical interest: A bibliographic review}",
      journal = {\planss},
         year = 2009,
        month = dec,
       volume = {57},
       number = {14-15},
        pages = {2053-2080},
          doi = {10.1016/j.pss.2009.09.011},
       adsurl = {https://ui.adsabs.harvard.edu/abs/2009P&SS...57.2053F}
}

@ARTICLE{dellorusso2016,
       author = {{Dello Russo}, Neil and {Kawakita}, Hideyo and {Vervack}, Ronald J. and {Weaver}, Harold A.},
        title = "{Emerging trends and a comet taxonomy based on the volatile chemistry measured in thirty comets with high-resolution infrared spectroscopy between 1997 and 2013}",
      journal = {\icarus},
         year = 2016,
        month = nov,
       volume = {278},
        pages = {301-332},
          doi = {10.1016/j.icarus.2016.05.039},
       adsurl = {https://ui.adsabs.harvard.edu/abs/2016Icar..278..301D}
}

@ARTICLE{Henault2025,
       author = {{H{\'e}nault}, E. and {Baklouti}, D. and {Brunetto}, R. and {Djouadi}, Z. and {Urso}, R.~G. and {Benoit-Lamaitrie}, P. and {Bour{\c{c}}ois}, J. and {Mivumbi}, O. and {Dalle Ore}, C.~M. and {Ricca}, A.},
        title = "{Methanol on red TNOs: A link between early composition and irradiation history}",
      journal = {\icarus},
         year = 2025,
        month = nov,
       volume = {441},
          eid = {116669},
        pages = {116669},
          doi = {10.1016/j.icarus.2025.116669},
       adsurl = {https://ui.adsabs.harvard.edu/abs/2025Icar..44116669H}
}

@ARTICLE{protopapa2020,
       author = {{Protopapa}, Silvia and {Olkin}, Cathy B. and {Grundy}, Will M. and {Li}, Jian-Yang and {Verbiscer}, Anne and {Cruikshank}, Dale P. and {Gautier}, Thomas and {Quirico}, Eric and {Cook}, Jason C. and {Reuter}, Dennis and {Howett}, Carly J.~A. and {Stern}, Alan and {Beyer}, Ross A. and {Porter}, Simon and {Young}, Leslie A. and {Weaver}, Hal A. and {Ennico}, Kim and {Dalle Ore}, Cristina M. and {Scipioni}, Francesca and {Singer}, Kelsi},
        title = "{Disk-resolved Photometric Properties of Pluto and the Coloring Materials across its Surface}",
      journal = {\aj},
         year = 2020,
        month = feb,
       volume = {159},
       number = {2},
          eid = {74},
        pages = {74},
          doi = {10.3847/1538-3881/ab5e82},
       adsurl = {https://ui.adsabs.harvard.edu/abs/2020AJ....159...74P}
}

@BOOK{bohren1983,
       author = {{Bohren}, Craig F. and {Huffman}, Donald R.},
        title = "{Absorption and scattering of light by small particles}",
         year = 1983,
       adsurl = {https://ui.adsabs.harvard.edu/abs/1983asls.book.....B}
}

@ARTICLE{verbiscer2022,
       author = {{Verbiscer}, Anne J. and {Helfenstein}, Paul and {Porter}, Simon B. and {Benecchi}, Susan D. and {Kavelaars}, J.~J. and {Lauer}, Tod R. and {Peng}, Jinghan and {Protopapa}, Silvia and {Spencer}, John R. and {Stern}, S. Alan and {Weaver}, Harold A. and {Buie}, Marc W. and {Buratti}, Bonnie J. and {Olkin}, Catherine B. and {Parker}, Joel and {Singer}, Kelsi N. and {Young}, Leslie A. and {New Horizons Science Team}},
        title = "{The Diverse Shapes of Dwarf Planet and Large KBO Phase Curves Observed from New Horizons}",
      journal = {\psj},
         year = 2022,
        month = apr,
       volume = {3},
       number = {4},
          eid = {95},
        pages = {95},
          doi = {10.3847/PSJ/ac63a6},
       adsurl = {https://ui.adsabs.harvard.edu/abs/2022PSJ.....3...95V}
}

@ARTICLE{mastrapa2009,
       author = {{Mastrapa}, R.~M. and {Sandford}, S.~A. and {Roush}, T.~L. and {Cruikshank}, D.~P. and {Dalle Ore}, C.~M.},
        title = "{Optical Constants of Amorphous and Crystalline H$_{2}$O-ice: 2.5-22 {\ensuremath{\mu}}m (4000-455 cm$^{-1}$) Optical Constants of H$_{2}$O-ice}",
      journal = {\apj},
         year = 2009,
        month = aug,
       volume = {701},
       number = {2},
        pages = {1347-1356},
          doi = {10.1088/0004-637X/701/2/1347},
       adsurl = {https://ui.adsabs.harvard.edu/abs/2009ApJ...701.1347M}
}

@INCOLLECTION{prialnik2004,
       author = {{Prialnik}, D. and {Benkhoff}, J. and {Podolak}, M.},
        title = "{Modeling the structure and activity of comet nuclei}",
    booktitle = {Comets II},
         year = 2004,
       editor = {{Festou}, Michel C. and {Keller}, H. Uwe and {Weaver}, Harold A.},
        pages = {359},
       adsurl = {https://ui.adsabs.harvard.edu/abs/2004come.book..359P}
}

@ARTICLE{hudson2021,
    author = {{Hudson}, Reggie L. and {Gerakines}, Perry A. and {Yarnall}, Yukiko Y. and {Coones}, Ryan T.},
    title = "{Infrared spectra and optical constants of astronomical ices: III. Propane, propylene, and propyne}",
    journal = {\icarus},
    year = 2021,
    month = jan,
    volume = {354},
    eid = {114033},
    pages = {114033},
    doi = {10.1016/j.icarus.2020.114033},
    adsurl = {https://ui.adsabs.harvard.edu/abs/2021Icar..35414033H}
}

@ARTICLE{hudson2014,
    author = {{Hudson}, R.~L. and {Ferrante}, R.~F. and {Moore}, M.~H.},
    title = "{Infrared spectra and optical constants of astronomical ices: I. Amorphous and crystalline acetylene}",
    journal = {\icarus},
    year = 2014,
    month = jan,
    volume = {228},
    pages = {276-287},
    doi = {10.1016/j.icarus.2013.08.029},
    adsurl = {https://ui.adsabs.harvard.edu/abs/2014Icar..228..276H}
}

@ARTICLE{fornasier2013,
       author = {{Fornasier}, S. and {Lellouch}, E. and {M{\"u}ller}, T. and {Santos-Sanz}, P. and {Panuzzo}, P. and {Kiss}, C. and {Lim}, T. and {Mommert}, M. and {Bockel{\'e}e-Morvan}, D. and {Vilenius}, E. and {Stansberry}, J. and {Tozzi}, G.~P. and {Mottola}, S. and {Delsanti}, A. and {Crovisier}, J. and {Duffard}, R. and {Henry}, F. and {Lacerda}, P. and {Barucci}, A. and {Gicquel}, A.},
        title = "{TNOs are Cool: A survey of the trans-Neptunian region. VIII. Combined Herschel PACS and SPIRE observations of nine bright targets at 70-500 {\ensuremath{\mu}}m}",
      journal = {\aap},
         year = 2013,
        month = jul,
       volume = {555},
          eid = {A15},
        pages = {A15},
          doi = {10.1051/0004-6361/201321329},
archivePrefix = {arXiv},
       eprint = {1305.0449},
 primaryClass = {astro-ph.EP},
       adsurl = {https://ui.adsabs.harvard.edu/abs/2013A&A...555A..15F}
}

@ARTICLE{khare1984,
       author = {{Khare}, B.~N. and {Sagan}, C. and {Arakawa}, E.~T. and {Suits}, F. and {Callcott}, T.~A. and {Williams}, M.~W.},
        title = "{Optical constants of organic tholins produced in a simulated Titanian atmosphere: From soft x-ray to microwave frequencies}",
      journal = {\icarus},
         year = 1984,
        month = oct,
       volume = {60},
       number = {1},
        pages = {127-137},
          doi = {10.1016/0019-1035(84)90142-8},
       adsurl = {https://ui.adsabs.harvard.edu/abs/1984Icar...60..127K}
}

@ARTICLE{hansen1997,
       author = {{Hansen}, Gary B.},
        title = "{The infrared absorption spectrum of carbon dioxide ice from 1.8 to 333 {\ensuremath{\mu}}m}",
      journal = {\jgr},
         year = 1997,
        month = sep,
       volume = {102},
       number = {E9},
        pages = {21569-21588},
          doi = {10.1029/97JE01875},
       adsurl = {https://ui.adsabs.harvard.edu/abs/1997JGR...10221569H}
}

@ARTICLE{protopapa2025,
       author = {{Protopapa}, Silvia and {Wong}, Ian and {Lellouch}, Emmanuel and {Johnson}, Perianne E. and {Grundy}, William M. and {Glein}, Christopher R. and {M{\"u}ller}, Thomas and {Kiss}, Csaba and {Emery}, Joshua P. and {Brunetto}, Rosario and {Holler}, Bryan J. and {Parker}, Alex H. and {Stansberry}, John A. and {Hammel}, Heidi B. and {Milam}, Stefanie N. and {Guilbert-Lepoutre}, Aur{\'e}lie and {Santos-Sanz}, Pablo and {Pinilla-Alonso}, Noem{\'\i}},
        title = "{JWST Detection of Hydrocarbon Ices and Methane Gas on Makemake}",
      journal = {\apjl},
         year = 2025,
        month = oct,
       volume = {991},
       number = {2},
          eid = {L34},
        pages = {L34},
          doi = {10.3847/2041-8213/adfe63},
archivePrefix = {arXiv},
       eprint = {2509.06772},
 primaryClass = {astro-ph.EP},
       adsurl = {https://ui.adsabs.harvard.edu/abs/2025ApJ...991L..34P}
}

@ARTICLE{snodgrass2025,
       author = {{Snodgrass}, Colin and {Holt}, Carrie E. and {Kelley}, Michael S.~P. and {Opitom}, Cyrielle and {Guilbert-Lepoutre}, Aur{\'e}lie and {Knight}, Matthew M. and {Kokotanekova}, Rosita and {Jehin}, Emmanuel and {Mazzotta Epifani}, Elena and {Migliorini}, Alessandra and {Tubiana}, Cecilia and {Micheli}, Marco and {Farnocchia}, Davide},
        title = "{First JWST spectrum of distant activity in long-period comet C/2024 E1 (Wierzchos)}",
      journal = {\mnras},
         year = 2025,
        month = jul,
       volume = {541},
       number = {1},
        pages = {L8-L13},
          doi = {10.1093/mnrasl/slaf046},
archivePrefix = {arXiv},
       eprint = {2503.14071},
 primaryClass = {astro-ph.EP},
       adsurl = {https://ui.adsabs.harvard.edu/abs/2025MNRAS.541L...8S}
}

@ARTICLE{jewitt2009,
       author = {{Jewitt}, David},
        title = "{The Active Centaurs}",
      journal = {\aj},
         year = 2009,
        month = may,
       volume = {137},
       number = {5},
        pages = {4296-4312},
          doi = {10.1088/0004-6256/137/5/4296},
archivePrefix = {arXiv},
       eprint = {0902.4687},
 primaryClass = {astro-ph.EP},
       adsurl = {https://ui.adsabs.harvard.edu/abs/2009AJ....137.4296J}
}

@INCOLLECTION{protopapa2021,
    author = {{Protopapa}, S. and {Cook}, J.~C. and {Grundy}, W.~M. and {Cruikshank}, D.~P. and {Dalle Ore}, C.~M. and {Beyer}, R.~A.},
    title = "{Surface Composition of Charon}",
    booktitle = {The Pluto System After New Horizons},
    year = 2021,
    editor = {{Stern}, S.~A. and {Moore}, J.~M. and {Grundy}, W.~M. and {Young}, L.~A. and {Binzel}, R.~P.},
    pages = {433-456},
    doi = {10.2458/azu_uapress_9780816540945-ch019},
    adsurl = {https://ui.adsabs.harvard.edu/abs/2021psnh.book..433P}
}

@ARTICLE{khare1993,
       author = {{Khare}, B.~N. and {Thompson}, W.~R. and {Cheng}, L. and {Chyba}, C. and {Sagan}, C. and {Arakawa}, E.~T. and {Meisse}, C. and {Tuminello}, P.~S.},
        title = "{Production and Optical Constants of Ice Tholin from Charged Particle Irradiation of (1:6) C $_{2}$H $_{6}$/H $_{2}$O at 77 K}",
      journal = {\icarus},
         year = 1993,
        month = jun,
       volume = {103},
       number = {2},
        pages = {290-300},
          doi = {10.1006/icar.1993.1071},
       adsurl = {https://ui.adsabs.harvard.edu/abs/1993Icar..103..290K}
}

@ARTICLE{harringtonpinto2023,
       author = {{Harrington Pinto}, O. and {Kelley}, M.~S.~P. and {Villanueva}, G.~L. and {Womack}, M. and {Faggi}, S. and {McKay}, A. and {DiSanti}, M.~A. and {Schambeau}, C. and {Fernandez}, Y. and {Bauer}, J. and {Feaga}, L. and {Wierzchos}, K.},
        title = "{First Detection of CO$_{2}$ Emission in a Centaur: JWST NIRSpec Observations of 39P/Oterma}",
      journal = {\psj},
         year = 2023,
        month = nov,
       volume = {4},
       number = {11},
          eid = {208},
        pages = {208},
          doi = {10.3847/PSJ/acf928},
archivePrefix = {arXiv},
       eprint = {2309.11486},
 primaryClass = {astro-ph.EP},
       adsurl = {https://ui.adsabs.harvard.edu/abs/2023PSJ.....4..208H}
}

@ARTICLE{faggi2024,
       author = {{Faggi}, Sara and {Villanueva}, Geronimo L. and {McKay}, Adam and {Harrington Pinto}, Olga and {Kelley}, Michael S.~P. and {Bockel{\'e}e-Morvan}, Dominique and {Womack}, Maria and {Schambeau}, Charles A. and {Feaga}, Lori and {DiSanti}, Michael A. and {Bauer}, James M. and {Biver}, Nicolas and {Wierzchos}, Kacper and {Fernandez}, Yanga R.},
        title = "{Heterogeneous outgassing regions identified on active centaur 29P/Schwassmann{\textendash}Wachmann 1}",
      journal = {Nature Astronomy},
         year = 2024,
        month = oct,
       volume = {8},
       number = {10},
        pages = {1237-1245},
          doi = {10.1038/s41550-024-02319-3},
       adsurl = {https://ui.adsabs.harvard.edu/abs/2024NatAs...8.1237F}
}

@ARTICLE{protopapa2024,
       author = {{Protopapa}, Silvia and {Raut}, Ujjwal and {Wong}, Ian and {Stansberry}, John and {Villanueva}, Geronimo L. and {Cook}, Jason and {Holler}, Bryan and {Grundy}, William M. and {Brunetto}, Rosario and {Cartwright}, Richard J. and {Mamo}, Bereket and {Emery}, Joshua P. and {Parker}, Alex H. and {Guilbert-Lepoutre}, Aurelie and {Pinilla-Alonso}, Noemi and {Milam}, Stefanie N. and {Hammel}, Heidi B.},
        title = "{Detection of carbon dioxide and hydrogen peroxide on the stratified surface of Charon with JWST}",
      journal = {Nature Communications},
         year = 2024,
        month = dec,
       volume = {15},
       number = {1},
          eid = {8247},
        pages = {8247},
          doi = {10.1038/s41467-024-51826-4},
       adsurl = {https://ui.adsabs.harvard.edu/abs/2024NatCo..15.8247P}
}

@ARTICLE{holler2016,
       author = {{Holler}, B.~J. and {Young}, L.~A. and {Grundy}, W.~M. and {Olkin}, C.~B.},
        title = "{On the surface composition of Triton's southern latitudes}",
      journal = {\icarus},
         year = 2016,
        month = mar,
       volume = {267},
        pages = {255-266},
          doi = {10.1016/j.icarus.2015.12.027},
archivePrefix = {arXiv},
       eprint = {1508.05924},
 primaryClass = {astro-ph.EP},
       adsurl = {https://ui.adsabs.harvard.edu/abs/2016Icar..267..255H}
}

@ARTICLE{grundy2010,
       author = {{Grundy}, W.~M. and {Young}, L.~A. and {Stansberry}, J.~A. and {Buie}, M.~W. and {Olkin}, C.~B. and {Young}, E.~F.},
        title = "{Near-infrared spectral monitoring of Triton with IRTF/SpeX II: Spatial distribution and evolution of ices}",
      journal = {\icarus},
         year = 2010,
        month = feb,
       volume = {205},
       number = {2},
        pages = {594-604},
          doi = {10.1016/j.icarus.2009.08.005},
archivePrefix = {arXiv},
       eprint = {0908.2623},
 primaryClass = {astro-ph.EP},
       adsurl = {https://ui.adsabs.harvard.edu/abs/2010Icar..205..594G}
}

@ARTICLE{quirico1999,
       author = {{Quirico}, Eric and {Dout{\'e}}, Sylvain and {Schmitt}, Bernard and {de Bergh}, Catherine and {Cruikshank}, Dale P. and {Owen}, Tobias C. and {Geballe}, Thomas R. and {Roush}, Ted L.},
        title = "{Composition, Physical State, and Distribution of Ices at the Surface of Triton}",
      journal = {\icarus},
         year = 1999,
        month = jun,
       volume = {139},
       number = {2},
        pages = {159-178},
          doi = {10.1006/icar.1999.6111},
       adsurl = {https://ui.adsabs.harvard.edu/abs/1999Icar..139..159Q}
}

@ARTICLE{cruikshank1993,
       author = {{Cruikshank}, D.~P. and {Roush}, T.~L. and {Owen}, T.~C. and {Geballe}, T.~R. and {de Bergh}, C. and {Schmitt}, B. and {Brown}, R.~H. and {Bartholomew}, M.~J.},
        title = "{Ices on the Surface of Triton}",
      journal = {Science},
         year = 1993,
        month = aug,
       volume = {261},
       number = {5122},
        pages = {742-745},
          doi = {10.1126/science.261.5122.742},
       adsurl = {https://ui.adsabs.harvard.edu/abs/1993Sci...261..742C}
}

@ARTICLE{soubiran2018,
       author = {{Soubiran}, C. and {Jasniewicz}, G. and {Chemin}, L. and {Zurbach}, C. and {Brouillet}, N. and {Panuzzo}, P. and {Sartoretti}, P. and {Katz}, D. and {Le Campion}, J.-F. and {Marchal}, O. and {Hestroffer}, D. and {Th{\'e}venin}, F. and {Crifo}, F. and {Udry}, S. and {Cropper}, M. and {Seabroke}, G. and {Viala}, Y. and {Benson}, K. and {Blomme}, R. and {Jean-Antoine}, A. and {Huckle}, H. and {Smith}, M. and {Baker}, S.~G. and {Damerdji}, Y. and {Dolding}, C. and {Fr{\'e}mat}, Y. and {Gosset}, E. and {Guerrier}, A. and {Guy}, L.~P. and {Haigron}, R. and {Jan{\ss}en}, K. and {Plum}, G. and {Fabre}, C. and {Lasne}, Y. and {Pailler}, F. and {Panem}, C. and {Riclet}, F. and {Royer}, F. and {Tauran}, G. and {Zwitter}, T. and {Gueguen}, A. and {Turon}, C.},
        title = "{Gaia Data Release 2. The catalogue of radial velocity standard stars}",
      journal = {\aap},
         year = 2018,
        month = aug,
       volume = {616},
          eid = {A7},
        pages = {A7},
          doi = {10.1051/0004-6361/201832795},
archivePrefix = {arXiv},
       eprint = {1804.09370},
 primaryClass = {astro-ph.GA},
       adsurl = {https://ui.adsabs.harvard.edu/abs/2018A&A...616A...7S}
}

@ARTICLE{boker2023,
       author = {{B{\"o}ker}, T. and {Beck}, T.~L. and {Birkmann}, S.~M. and {Giardino}, G. and {Keyes}, C. and {Kumari}, N. and {Muzerolle}, J. and {Rawle}, T. and {Zeidler}, P. and {Abul-Huda}, Y. and {Alves de Oliveira}, C. and {Arribas}, S. and {Bechtold}, K. and {Bhatawdekar}, R. and {Bonaventura}, N. and {Bunker}, A.~J. and {Cameron}, A.~J. and {Carniani}, S. and {Charlot}, S. and {Curti}, M. and {Espinoza}, N. and {Ferruit}, P. and {Franx}, M. and {Jakobsen}, P. and {Karakla}, D. and {L{\'o}pez-Caniego}, M. and {L{\"u}tzgendorf}, N. and {Maiolino}, R. and {Manjavacas}, E. and {Marston}, A.~P. and {Moseley}, S.~H. and {Ogle}, P. and {Perna}, M. and {Pe{\~n}a-Guerrero}, M. and {Pirzkal}, N. and {Plesha}, R. and {Proffitt}, C.~R. and {Rauscher}, B.~J. and {Rix}, H. -W. and {Rodr{\'\i}guez del Pino}, B. and {Rustamkulov}, Z. and {Sabbi}, E. and {Sing}, D.~K. and {Sirianni}, M. and {te Plate}, M. and {{\'U}beda}, L. and {Wahlgren}, G.~M. and {Wislowski}, E. and {Wu}, R. and {Willott}, Chris J.},
        title = "{In-orbit Performance of the Near-infrared Spectrograph NIRSpec on the James Webb Space Telescope}",
      journal = {\pasp},
         year = 2023,
        month = mar,
       volume = {135},
       number = {1045},
          eid = {038001},
        pages = {038001},
          doi = {10.1088/1538-3873/acb846},
archivePrefix = {arXiv},
       eprint = {2301.13766},
 primaryClass = {astro-ph.IM},
       adsurl = {https://ui.adsabs.harvard.edu/abs/2023PASP..135c8001B}
}

@ARTICLE{jakobsen2022,
       author = {{Jakobsen}, P. and {Ferruit}, P. and {Alves de Oliveira}, C. and {Arribas}, S. and {Bagnasco}, G. and {Barho}, R. and {Beck}, T.~L. and {Birkmann}, S. and {B{\"o}ker}, T. and {Bunker}, A.~J. and {Charlot}, S. and {de Jong}, P. and {de Marchi}, G. and {Ehrenwinkler}, R. and {Falcolini}, M. and {Fels}, R. and {Franx}, M. and {Franz}, D. and {Funke}, M. and {Giardino}, G. and {Gnata}, X. and {Holota}, W. and {Honnen}, K. and {Jensen}, P.~L. and {Jentsch}, M. and {Johnson}, T. and {Jollet}, D. and {Karl}, H. and {Kling}, G. and {K{\"o}hler}, J. and {Kolm}, M. -G. and {Kumari}, N. and {Lander}, M.~E. and {Lemke}, R. and {L{\'o}pez-Caniego}, M. and {L{\"u}tzgendorf}, N. and {Maiolino}, R. and {Manjavacas}, E. and {Marston}, A. and {Maschmann}, M. and {Maurer}, R. and {Messerschmidt}, B. and {Moseley}, S.~H. and {Mosner}, P. and {Mott}, D.~B. and {Muzerolle}, J. and {Pirzkal}, N. and {Pittet}, J. -F. and {Plitzke}, A. and {Posselt}, W. and {Rapp}, B. and {Rauscher}, B.~J. and {Rawle}, T. and {Rix}, H. -W. and {R{\"o}del}, A. and {Rumler}, P. and {Sabbi}, E. and {Salvignol}, J. -C. and {Schmid}, T. and {Sirianni}, M. and {Smith}, C. and {Strada}, P. and {te Plate}, M. and {Valenti}, J. and {Wettemann}, T. and {Wiehe}, T. and {Wiesmayer}, M. and {Willott}, C.~J. and {Wright}, R. and {Zeidler}, P. and {Zincke}, C.},
        title = "{The Near-Infrared Spectrograph (NIRSpec) on the James Webb Space Telescope. I. Overview of the instrument and its capabilities}",
      journal = {\aap},
         year = 2022,
        month = may,
       volume = {661},
          eid = {A80},
        pages = {A80},
          doi = {10.1051/0004-6361/202142663},
archivePrefix = {arXiv},
       eprint = {2202.03305},
 primaryClass = {astro-ph.IM},
       adsurl = {https://ui.adsabs.harvard.edu/abs/2022A&A...661A..80J}
}

@ARTICLE{emery2024,
       author = {{Emery}, J.~P. and {Wong}, I. and {Brunetto}, R. and {Cook}, J.~C. and {Pinilla-Alonso}, N. and {Stansberry}, J.~A. and {Holler}, B.~J. and {Grundy}, W.~M. and {Protopapa}, S. and {Souza-Feliciano}, A.~C. and {Fern{\'a}ndez-Valenzuela}, E. and {Lunine}, J.~I. and {Hines}, D.~C.},
        title = "{A tale of 3 dwarf planets: Ices and organics on Sedna, Gonggong, and Quaoar from JWST spectroscopy}",
      journal = {\icarus},
         year = 2024,
        month = may,
       volume = {414},
          eid = {116017},
        pages = {116017},
          doi = {10.1016/j.icarus.2024.116017},
archivePrefix = {arXiv},
       eprint = {2309.15230},
 primaryClass = {astro-ph.EP},
       adsurl = {https://ui.adsabs.harvard.edu/abs/2024Icar..41416017E}
}

@INPROCEEDINGS{moseley2010,
       author = {{Moseley}, S.~H. and {Arendt}, Richard G. and {Fixsen}, D.~J. and {Lindler}, Don and {Loose}, Markus and {Rauscher}, Bernard J.},
        title = "{Reducing the read noise of H2RG detector arrays: eliminating correlated noise with efficient use of reference signals}",
    booktitle = {High Energy, Optical, and Infrared Detectors for Astronomy IV},
         year = 2010,
       editor = {{Holland}, Andrew D. and {Dorn}, David A.},
       series = {Society of Photo-Optical Instrumentation Engineers (SPIE) Conference Series},
       volume = {7742},
        month = jul,
          eid = {77421B},
        pages = {77421B},
          doi = {10.1117/12.866773},
       adsurl = {https://ui.adsabs.harvard.edu/abs/2010SPIE.7742E..1BM}
}

@INPROCEEDINGS{rauscher2012,
       author = {{Rauscher}, Bernard J. and {Arendt}, Richard G. and {Fixsen}, D.~J. and {Lander}, Matthew and {Lindler}, Don and {Loose}, Markus and {Moseley}, S.~H. and {Wilson}, Donna V. and {Xenophontos}, Christos},
        title = "{Reducing the read noise of HAWAII-2RG detector systems with improved reference sampling and subtraction (IRS$^{2}$)}",
    booktitle = {High Energy, Optical, and Infrared Detectors for Astronomy V},
         year = 2012,
       editor = {{Holland}, Andrew D. and {Beletic}, James W.},
       series = {Society of Photo-Optical Instrumentation Engineers (SPIE) Conference Series},
       volume = {8453},
        month = jul,
          eid = {84531F},
        pages = {84531F},
          doi = {10.1117/12.926089},
       adsurl = {https://ui.adsabs.harvard.edu/abs/2012SPIE.8453E..1FR}
}

@ARTICLE{holler2025,
       author = {{Holler}, Bryan J. and {Brunetto}, Rosario and {Cruikshank}, Dale P. and {Cryan}, Sasha and {Guilbert-Lepoutre}, Aurelie and {Harvison}, Brittany and {Licandro}, Javier and {McClure}, Lucas T. and {M{\"u}ller}, Thomas G. and {Peixinho}, Nuno and {Pinilla-Alonso}, Noem{\'\i} and {Stansberry}, John A. and {Belyakov}, Matthew and {Benecchi}, Susan and {Brown}, Michael E. and {Cartwright}, Richard J. and {Collyer}, Cameron and {de Pr{\'a}}, M{\'a}rio N. and {Fraser}, Wesley C. and {Markwardt}, Larissa and {Melendy}, J.~J. and {Noonan}, John W. and {Protopapa}, Silvia and {Proudfoot}, Benjamin and {Sharkey}, Benjamin N.~L. and {Souza Feliciano}, Ana Carolina and {Verbiscer}, Anne J. and {Wong}, Ian and {Young}, Leslie A.},
        title = "{A Descriptive Taxonomic Nomenclature for Intermediate-sized Trans-Neptunian Object Spectra}",
      journal = {RNAAS},
         year = 2025,
        month = sep,
       volume = {9},
       number = {9},
          eid = {241},
        pages = {241},
          doi = {10.3847/2515-5172/ae03a2},
       adsurl = {https://ui.adsabs.harvard.edu/abs/2025RNAAS...9..241H}
}

@INPROCEEDINGS{delsemme1982,
       author = {{Delsemme}, A.~H.},
        title = "{Chemical composition of cometary nuclei}",
    booktitle = {IAU Colloquium 61: Comet Discoveries, Statistics, and Observational Selection},
         year = 1982,
       editor = {{Wilkening}, L.~L.},
        month = jan,
        pages = {85-130},
       adsurl = {https://ui.adsabs.harvard.edu/abs/1982come.coll...85D}
}

@misc{jwst,
       author = {{Bushouse}, Howard and {Eisenhamer}, Jonathan and {Dencheva}, Nadia and {Davies}, James and {Greenfield}, Perry and {Morrison}, Jane and {Hodge}, Phil and {Simon}, Bernie and {Grumm}, David and {Droettboom}, Michael and {Slavich}, Edward and {Sosey}, Megan and {Pauly}, Tyler and {Miller}, Todd and {Jedrzejewski}, Robert and {Hack}, Warren and {Davis}, David and {Crawford}, Steven and {Law}, David and {Gordon}, Karl and {Regan}, Michael and {Cara}, Mihai and {MacDonald}, Ken and {Bradley}, Larry and {Shanahan}, Clare and {Jamieson}, William and {Teodoro}, Mairan and {Williams}, Thomas and {Pena-Guerrero}, Maria},
        title = {{JWST Calibration Pipeline v1.13.4}},
    journal = {Zenodo},
         year = 2024,
        month = jan,
          eid = {10.5281/zenodo.10569856},
          doi = {10.5281/zenodo.10569856},
      version = {v1.13.4,},
    publisher = {Zenodo},
       adsurl = {https://ui.adsabs.harvard.edu/abs/2023zndo...6984365B}
}

@ARTICLE{rauscher2024,
       author = {{Rauscher}, Bernard J.},
        title = "{NSClean: An Algorithm for Removing Correlated Noise from JWST NIRSpec Images}",
      journal = {\pasp},
         year = 2024,
        month = jan,
       volume = {136},
       number = {1},
          eid = {015001},
        pages = {015001},
          doi = {10.1088/1538-3873/ad1b36},
archivePrefix = {arXiv},
       eprint = {2306.03250},
 primaryClass = {astro-ph.IM},
       adsurl = {https://ui.adsabs.harvard.edu/abs/2024PASP..136a5001R}
}

@ARTICLE{pinillaalonso2025,
       author = {{Pinilla-Alonso}, Noem{\'\i} and {Brunetto}, Rosario and {De Pr{\'a}}, M{\'a}rio N. and {Holler}, Bryan J. and {H{\'e}nault}, Elsa and {Feliciano}, Ana Carolina de Souza and {Lorenzi}, Vania and {Pendleton}, Yvonne J. and {Cruikshank}, Dale P. and {M{\"u}ller}, Thomas G. and {Stansberry}, John A. and {Emery}, Joshua P. and {Schambeau}, Charles A. and {Licandro}, Javier and {Harvison}, Brittany and {McClure}, Lucas and {Guilbert-Lepoutre}, Aur{\'e}lie and {Peixinho}, Nuno and {Bannister}, Michele T. and {Wong}, Ian},
        title = "{A JWST/DiSCo-TNOs portrait of the primordial Solar System through its trans-Neptunian objects}",
      journal = {NatAs},
         year = 2025,
        month = feb,
       volume = {9},
        pages = {230-244},
          doi = {10.1038/s41550-024-02433-2},
       adsurl = {https://ui.adsabs.harvard.edu/abs/2025NatAs...9..230P}
}

@ARTICLE{grundy2024,
       author = {{Grundy}, W.~M. and {Wong}, I. and {Glein}, C.~R. and {Protopapa}, S. and {Holler}, B.~J. and {Cook}, J.~C. and {Stansberry}, J.~A. and {Lunine}, J.~I. and {Parker}, A.~H. and {Hammel}, H.~B. and {Milam}, S.~N. and {Brunetto}, R. and {Pinilla-Alonso}, N. and {de Souza Feliciano}, A.~C. and {Emery}, J.~P. and {Licandro}, J.},
        title = "{Measurement of D/H and $^{13}$C/$^{12}$C ratios in methane ice on Eris and Makemake: Evidence for internal activity}",
      journal = {\icarus},
         year = 2024,
        month = mar,
       volume = {411},
          eid = {115923},
        pages = {115923},
          doi = {10.1016/j.icarus.2023.115923},
archivePrefix = {arXiv},
       eprint = {2309.05085},
 primaryClass = {astro-ph.EP},
       adsurl = {https://ui.adsabs.harvard.edu/abs/2024Icar..41115923G}
}

@ARTICLE{wong2024,
       author = {{Wong}, Ian and {Brown}, Michael E. and {Emery}, Joshua P. and {Binzel}, Richard P. and {Grundy}, William M. and {Marchi}, Simone and {Martin}, Audrey C. and {Noll}, Keith S. and {Sunshine}, Jessica M.},
        title = "{JWST Near-infrared Spectroscopy of the Lucy Jupiter Trojan Flyby Targets: Evidence for OH Absorption, Aliphatic Organics, and CO$_{2}$}",
      journal = {\psj},
         year = 2024,
        month = apr,
       volume = {5},
       number = {4},
          eid = {87},
        pages = {87},
          doi = {10.3847/PSJ/ad2fc3},
archivePrefix = {arXiv},
       eprint = {2311.11531},
 primaryClass = {astro-ph.EP},
       adsurl = {https://ui.adsabs.harvard.edu/abs/2024PSJ.....5...87W}
}

@ARTICLE{villanueva2011,
       author = {{Villanueva}, G.~L. and {Mumma}, M.~J. and {DiSanti}, M.~A. and {Bonev}, B.~P. and {Gibb}, E.~L. and {Magee-Sauer}, K. and {Blake}, G.~A. and {Salyk}, C.},
        title = "{The molecular composition of Comet C/2007 W1 (Boattini): Evidence of a peculiar outgassing and a rich chemistry}",
      journal = {\icarus},
         year = 2011,
        month = nov,
       volume = {216},
       number = {1},
        pages = {227-240},
          doi = {10.1016/j.icarus.2011.08.024},
       adsurl = {https://ui.adsabs.harvard.edu/abs/2011Icar..216..227V}
}

\bmhead{Acknowledgements}
This work is based on observations made with the NASA/ESA/CSA James Webb Space Telescope. The data were obtained from the Mikulski Archive for Space Telescopes at the Space Telescope Science Institute, which is operated by the Association of Universities for Research in Astronomy, Inc., under NASA contract NAS 5-03127 for JWST. These observations are associated with Program \#4621. R.B. and A.G.-L. gratefully acknowledge support from CNES (France) as part of their contributions to the JWST mission. N.P.-A. acknowledges the Ministry of Science, Innovation, and Universities (MICIU) in Spain and the State Agency for Research (AEI) for funding through the ATRAE program, project ATR2023-145683.

\begin{figure*}[ht]%
\centering
\includegraphics[width=\textwidth]{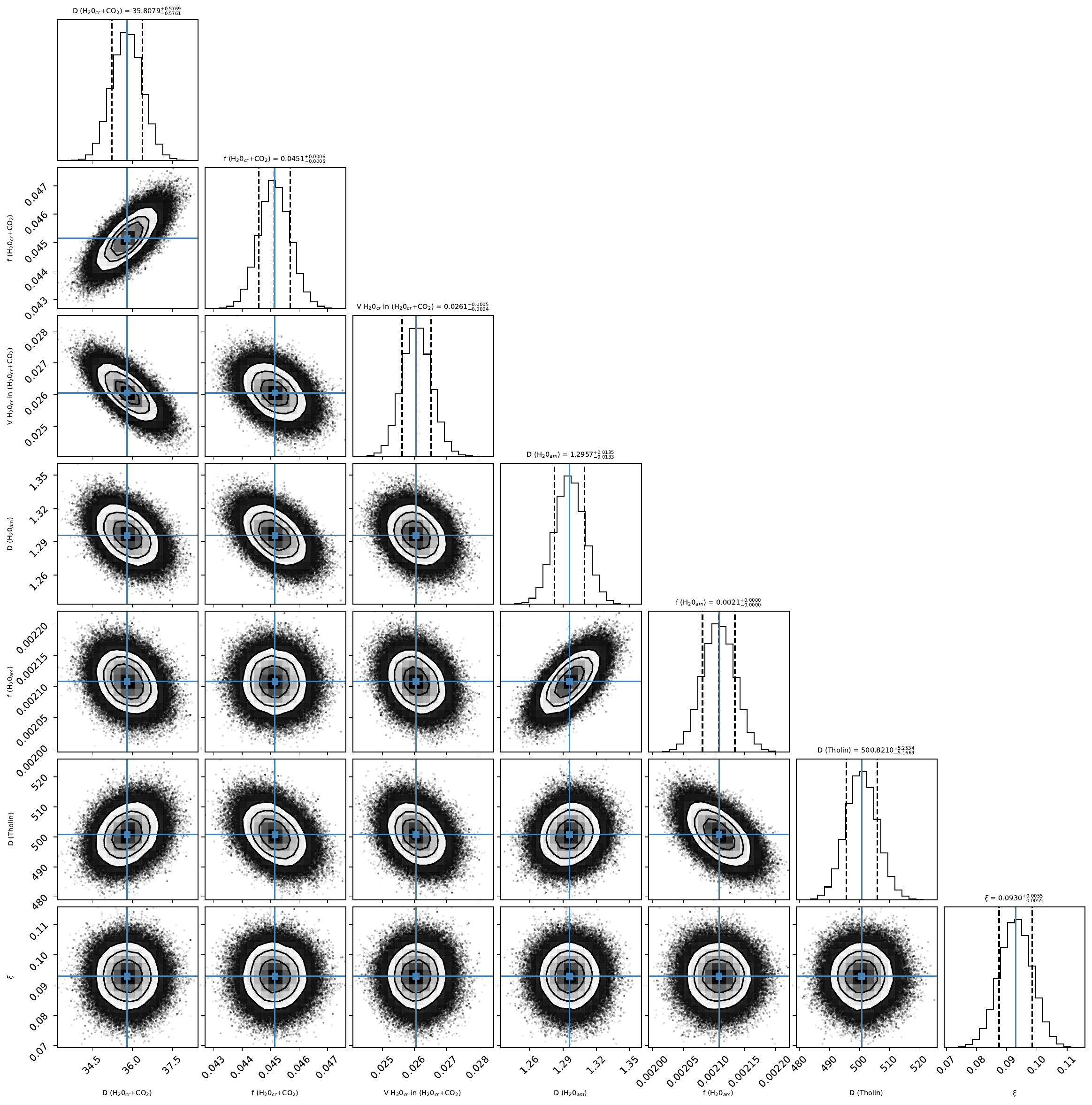}
\caption{\textbf{Posterior distributions of the free parameters from the radiative transfer modeling of Chiron's reflectance spectrum.} Corner plot showing the marginalized 1D and joint 2D posterior distributions for all free model parameters. Contours indicate the $1\sigma$, $2\sigma$, and $3\sigma$ regions. The median values and $1\sigma$ uncertainties are reported above each column. Cyan markers denote the best-fit parameter values obtained from the Levenberg--Marquardt least-squares minimization. See Methods for a description of the model parameters.}\label{fig:sm3}
\end{figure*}

\end{document}